\documentclass[prb,preprint,amsmath,amssymb]{revtex4}
\usepackage{graphicx}
\begin{document}
\author{Kevin Leung}
\affiliation{Sandia National Laboratories, MS 1415,
Albuquerque, NM 87185\\
$^*${\tt kleung@sandia.gov}}
\date{\today}
\title{Two-electron reduction of ethylene carbonate:
a quantum chemistry re-examination of mechanisms}

\input epsf
 
\begin{abstract}
 
Passivating solid-electrolyte interphase (SEI) films arising from
electrolyte decomposition on low-voltage lithium ion battery anode
surfaces are critical for battery operations.  We review the recent
theoretical literature on electrolyte decomposition and emphasize
the modeling work on two-electron reduction of ethylene carbonate
(EC, a key battery organic solvent).  One of the two-electron
pathways, which releases CO gas, is re-examined using simple quantum
chemistry calculations.  Excess electrons are shown to preferentially
attack EC in the order (broken EC$^-$) $>$ (intact EC$^-$) $>$ EC.
This confirms the viability of two electron processes and emphasizes
that they need to be considered when interpreting SEI experiments.
An estimate of the crossover between one- and two-electron regimes
under a homogeneous reaction zone approximation is proposed.
 
\end{abstract}
 
\maketitle
 
\section{Introduction}

\begin{table}\centering
\begin{tabular}{||l|l||} \hline
EC & charge neutral ethylene carbonate \\
c-EC$^-$ & intact ethylene carbonate radical anion \\
o-EC$^-$ & ring-opened ethylene carbonate radical anion \\
EDC & ethylene dicarbonate \\
BDC & butylene dicarbonate \\
$k_1$ & bimolecular EC$^-$ recombination rate to form BDC\\
$k_2$ & unimolecular EC$^{2-}$ decay rate \\
$k_3$ & unimolecular EC$^{-}$ ring-opening rate (C$_{\rm E}$-O$_{\rm E}$ bond) \\
$k_e$ & rate of electron tunneling to EC \\
$k_e'$ & rate of electron tunneling to EC$^-$ \\ \hline
\end{tabular}
\label{table0} \noindent
\end{table}

Solid electrolyte interphase (SEI) films on low voltage anode surfaces
(e.g., graphite, Li metal, Si) are critical for lithium ion battery
operations.\cite{book2,book,review,novak_review,intro1}  They arise from
electrochemical reduction and subsequent breakdown of the organic
solvent-based electrolyte which is metastable under battery charging voltage.
Once formed, the SEI hinders electron tunneling from the anode and prevents
further electrolyte decomposition while still permitting Li$^+$ ions to diffuse
between the electrolyte and the anode.  The electrolyte and electrode have
to be matched to produce stable SEI films.  For example, ethylene carbonate
(EC) is essential for widely used graphitic anodes.

Substantial experimental work has been performed to study the SEI structure
and chemical composition, which is extremely complex and heterogeneous.  The
gases released during the first charging cycle, when the SEI is largely
created, have also been analyzed.\cite{representative,gewirth,leifer,onuki,shin,ota1,ota2,yoshida,marom,tarascon,sasaki,edstrom,ogumi,endo,edc,novak1998,intro1}
Despite this, SEI formation mechanisms at the atomic lengthscale are difficult
to elucidate by purely experimental means, and significant uncertainties
remain.  With some exceptions, proposed mechanisms have been indirectly
inferred from SEI chemical composition and gas product distribution.  Such
analysis can be hampered by further reactions of initial electrolyte breakdown
products\cite{ogumi} and even sample preparation procedures during ex-situ
measurements.\cite{edstrom}  Battery material surfaces are not clean
or homogeneous, and differences in synthetic/experimental conditions likely 
contribute to SEI variations reported in different laboratories.  For example,
there are significant differences in the amount of CO gas
reported.\cite{yoshida,onuki,ota1,ota2,shin,novak1998,intro1}
As will be discussed, CO release is the signature of a key SEI formation
mechanism.

Electronic structure theoretical methods are well suited to elucidating
the initial electrolyte breakdown reaction pathways on both anodes and cathodes.\cite{bal01,bal02,bal02a,han2004,vollmer,pccp,bal11,brown,bedrov,ald,zhang01,johansson,tasaki,borodin}
Theoretical calculations are generally conducted under idealized conditions,
and at least at present they are not meant to reproduce the full range of
complex experimental conditions (e.g., electrode surface specificity,
impurities, state of charge, inhomogeneities or ``hotspots,'' interference
between cathodic and anodic processes) or the final SEI product distributions
(which may arise from multistep reactions).  To some extent, experiments and
theory are
complementary.  The greatest strength of electronic structure calculations is
arguably their ability to predict accurate reaction thermodynamics and barriers,
which are difficult to measure under the complex battery experimental
settings and the highly-driven, out-of-equilibrium conditions during
initial SEI growth.  Theory can help explain {\it why} certain reactions
are favorable, critically interrogate electrolyte degradation mechanisms
proposed in the literature (keeping in mind the more complex conditions
in experiments), and potentially uncover kinetically and thermodynamically
favorable reaction pathways that may become important under novel conditions
and can be taken advantage of.  The fundamental principles and insights
discovered through modeling can therefore contribute to the design of
``artificial SEI'' via new additives\cite{review} or electrode
coatings.\cite{ald}

Balbuena and coworkers pioneered the modeling of SEI formation and
decomposition of EC and other organic solvents.\cite{bal01,bal02,bal02a}
They applied DFT calculations of clusters of solvent molecules with
a Li$^+$, in the presence of an excess electron but without explicit
electrodes.  In recent years, there have been new developments
in modeling of SEI formation using more elaborate but costly computational
methods.  One recent work\cite{bedrov} may be interpreted as a major challenge
to a widely accepted one-electron electrolyte decomposition mechanism.  As that
mechanism was thought to yield ethylene dicarbonate (EDC), identified to
be a main SEI component,\cite{edc} this might be considered a ``crisis''
in SEI studies.  At the same time, DFT-based {\it ab initio} molecular
dynamics (AIMD) simulations of the initial stages of SEI growth at explicit
electrode-electrolyte interfaces have emphasized fast, two-electron attacks
on organic solvent molecules.  These simulations,\cite{pccp,bal11,ald} which
do not impose {\it a priori} reaction pathways, reproduce a mechanism and
CO/intermediate products that have been proposed and reported in the
literature\cite{yoshida,onuki,ota1,ota2,shin,marom} but are perhaps not
widely known.  

This paper takes a step back and applies easily reproduced cluster-based
calculations to integrate two-electron mechanisms, demonstrated in AIMD
simulations,\cite{pccp,bal11,ald} back into the original, elegant Balbuena
framework.\cite{bal01} We focus on intrinsic, two excess electron-induced
EC decomposition mechanisms.  Two-electron processes are
acknowledged\cite{book2,book,review} but have mostly been associated with
CO$_3^{2-}$ formation.  We show that the alternate, CO-releasing pathway is
less energetically favorable but exhibits a much lower intrinsic barrier.
We also point out the need to consider 2-$e^-$ pathways when explaining novel
experimental observations, including oligomerization.  Polymeric species are
present in the SEI,\cite{review} but so far the proposed mechanisms all
invoke one-electron pathways.\cite{gewirth,tarascon,yoshida,ogumi} 

By ``intrinsic'' mechanisms we mean excess $e^-$-induced reactions in the
bulk liquid region, with EC coordinated to Li$^+$ but not in the vicinity of
electrodes or counterions such as PF$_6^-$ or even cosolvents like dimethyl
carbonate (DMC), a linear alkyl carbonate.  The simplicity and clarity this
provides come at the price of ignoring important details.  Linear carbonates,
when reduced, generates alkoxide anions (RO$^-$) that can initiate nucleophilic
attack on EC.\cite{tarascon,sasaki}  PF$_6^-$ decomposition products are
part of SEI films.  It is also been shown to take part in EC oxidation near
the cathode.\cite{borodin}  SEI formation is clearly affected
by the electrode used.  For example, solvent intercalation between graphite
sheets\cite{intercal} competes with electrolyte decomposition on graphite
edges,\cite{markle} and Si-F covalent bonds in the SEI are unique to silicon
anodes.\cite{ckchan} While these electrode-specific effects are undoubtedly
present, the extent to which they are critical for SEI passivation has not
been completely estabilished.  Theoretical studies of electrode/electrolylte
interfaces constitute a topic unto itself and will be the subject of future
reviews.

Finally, to establish the importance of two electron processes, the $e^-$
tunneling rates associated with the first and second $e^-$ addition to EC
must be compared.  Computing electron transfer rate was first introduced
in the context of SEI formation in Ref.~\onlinecite{ald}. To that end, we
show that adding a second electron to EC is favored by both thermodynamic
(via the reduction potential, $\Phi$) and kinetic ($\Phi$ plus reorganization
energy, $\lambda$) considerations.  It is incorrect to consider tunneling
from the electrode as a pure wavepacket barrier-crossing problem, as is
the case in solid state devices.  Instead, the properties of the target
molecule must be included.  We will report a speculative, but to our
knowledge the first, estimate of the cross-over between the
1- and 2-$e^-$ regimes which yield different SEI products.

\section{Methods}

A few simple calculations are performed to exemplify 2-$e^-$ EC deomposition
pathways.  The M\"{o}ller-Plesset second order perturbation (MP2) method is
used to avoid ambiguities that may arise from using different DFT functionals.
Han and Lee\cite{han2004} have shown that MP2 reaction exothermicities
and barriers associated with EC$^-$ decomposition are comparable to highly
accurate CCSD(T) predictions.  The gaussian (g09) suite of programs\cite{g09}
and the ``SMD'' dielectric continuum approximation are applied,\cite{smd}
with $\epsilon$ set to the low frequency (40, approximated for the
entire battery electrolyte, not just liquid EC) or high frequency value
(2.62).\cite{ald}  Geometry optimization is performed using the
{\tt 6-31+G(d,p)} basis set.  Single point energies are reported at the
MP2/{\tt 6-311++G(3df,2pd)} level of theory.  Vibrational frequencies are
calculated for  stable structures and transition states.  This yields zero
point energies and thermal corrections which are included unless otherwise
noted.  Hybrid DFT functionals yield results similar to MP2, but for
simplicity those are not discussed.

[(EC)$_3$Li]$^{n-}$, isolated EC$^{m-}$, and EC$^{m-}$:Li$^+$ clusters are
considered.  Previous AIMD simulations have shown that Li$^+$ is coordinated
to slightly more than 3~EC molecules in EC liquid if one of the EC has an
excess electron.\cite{pccp}  A few AIMD simulations based on the PBE
functional\cite{pbe} are reported to complement MP2 studies.  Please
refer to Ref.~\onlinecite{pccp} for AIMD simulation details.

\section{Review of proposed mechanisms and modeling efforts}

\begin{figure}
\centerline{\hbox{ (a) \includegraphics[width=1.40in]{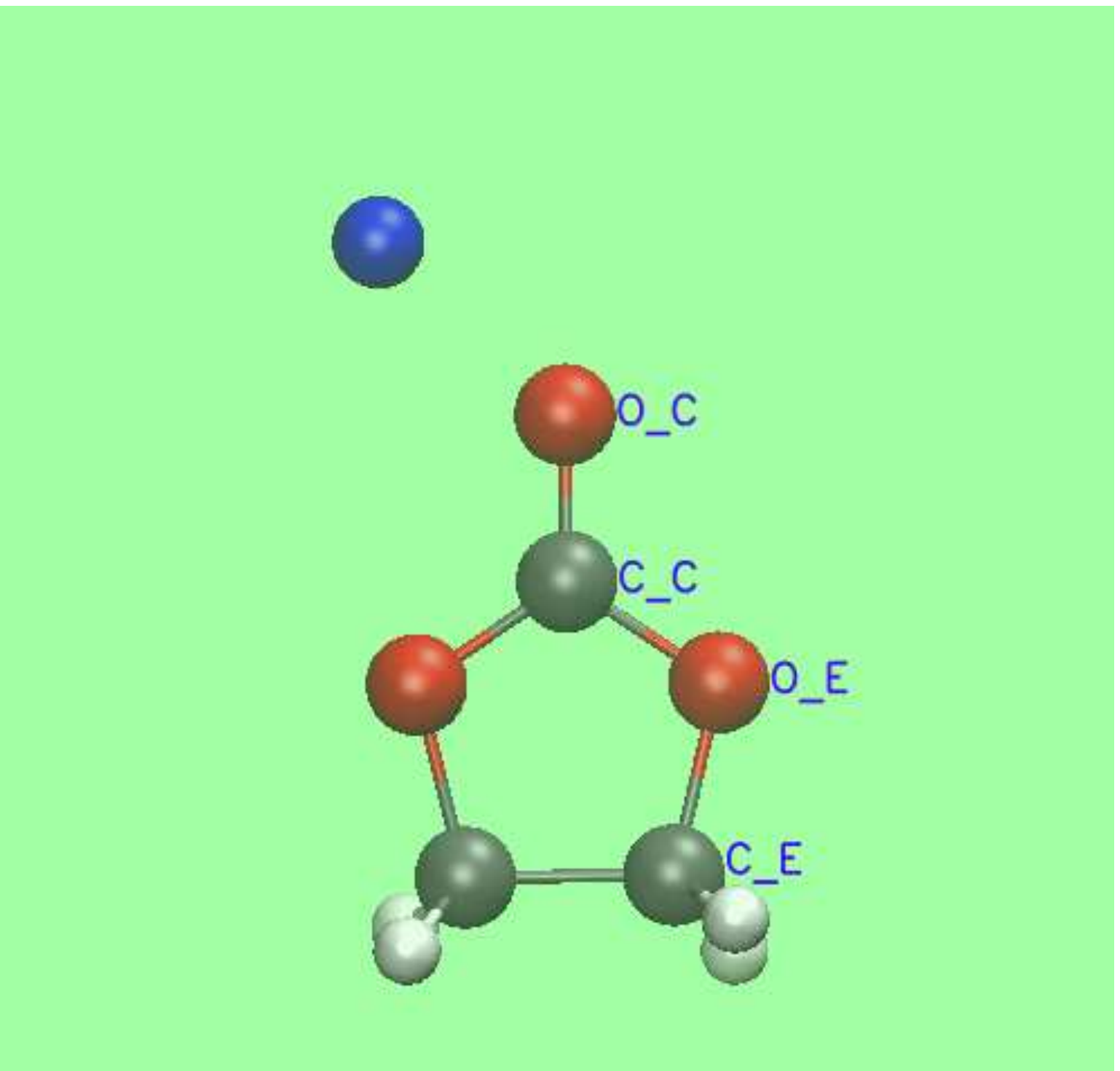}
		   (b) \includegraphics[width=1.40in]{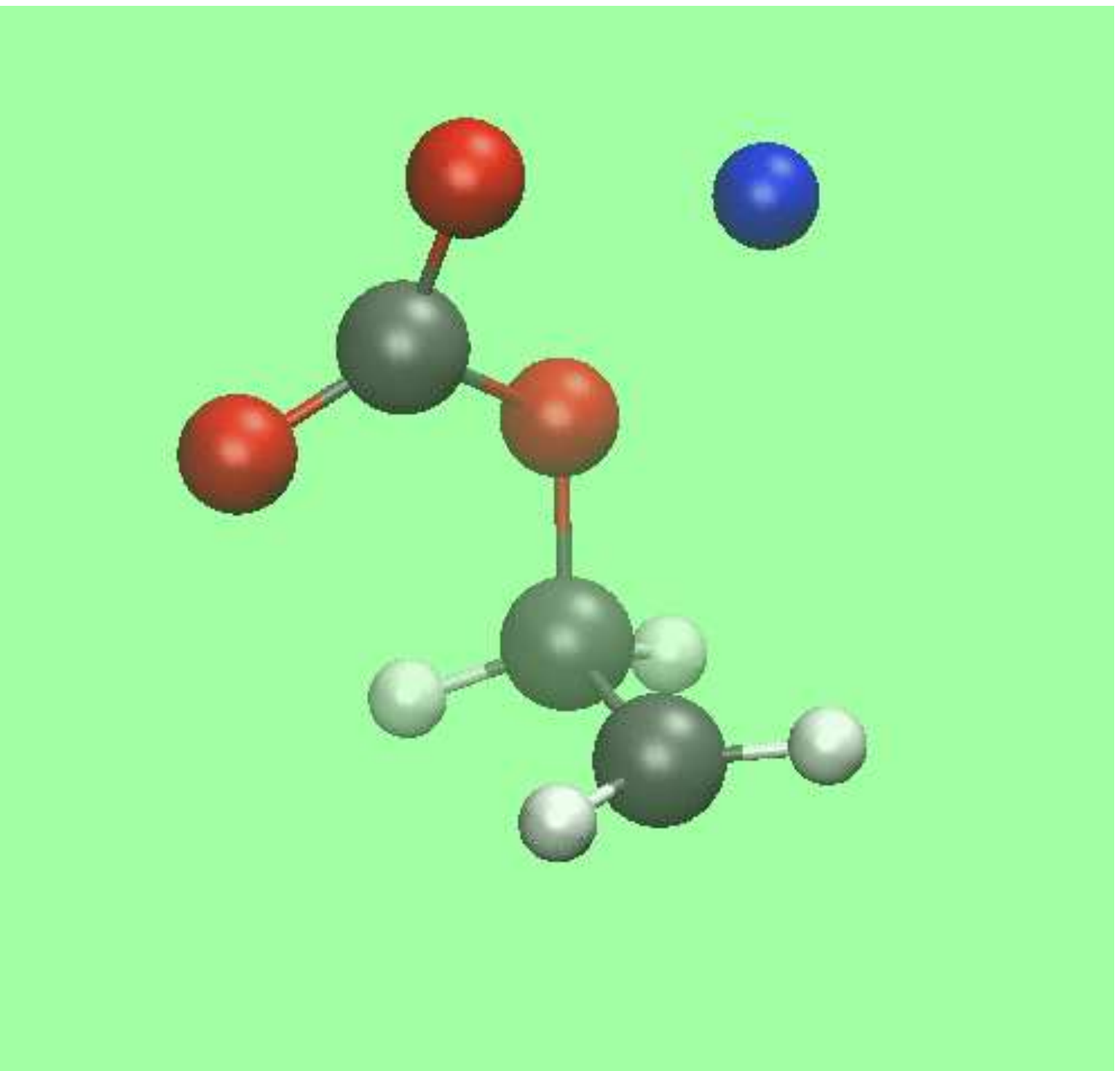}
		   (c) \includegraphics[width=1.40in]{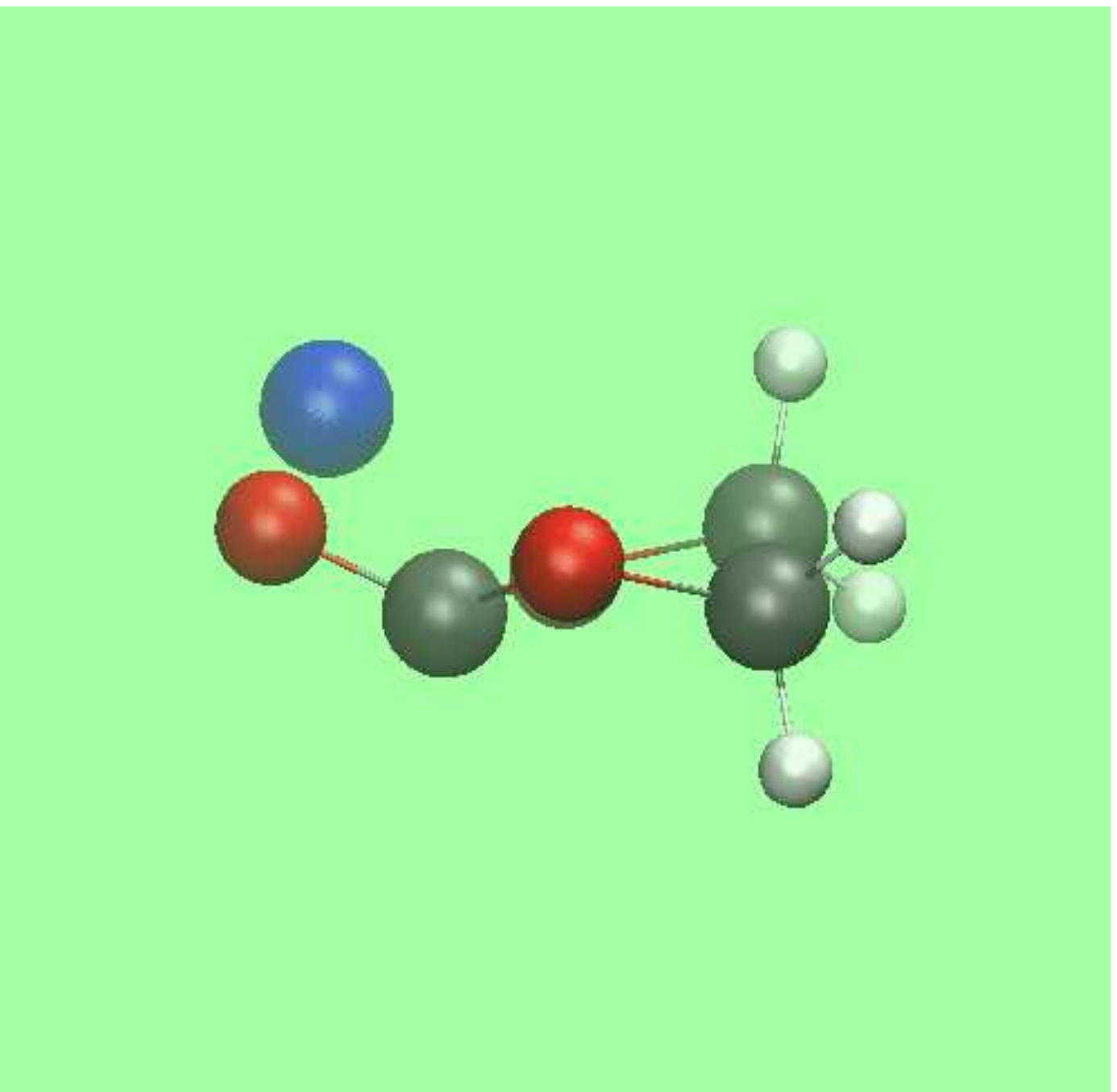} }}
\centerline{\hbox{ (d) \includegraphics[width=1.40in]{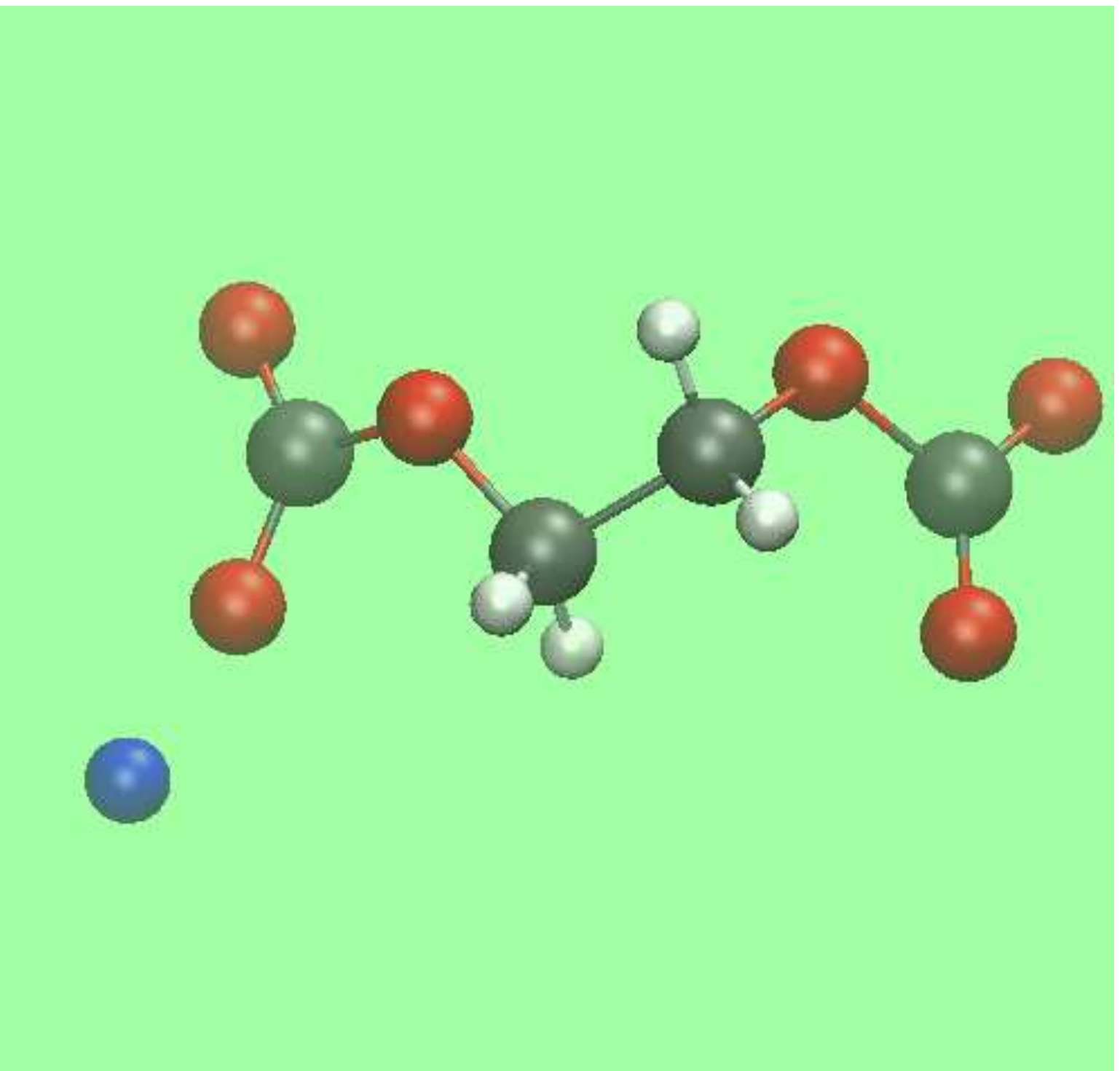}
		   (e) \includegraphics[width=1.40in]{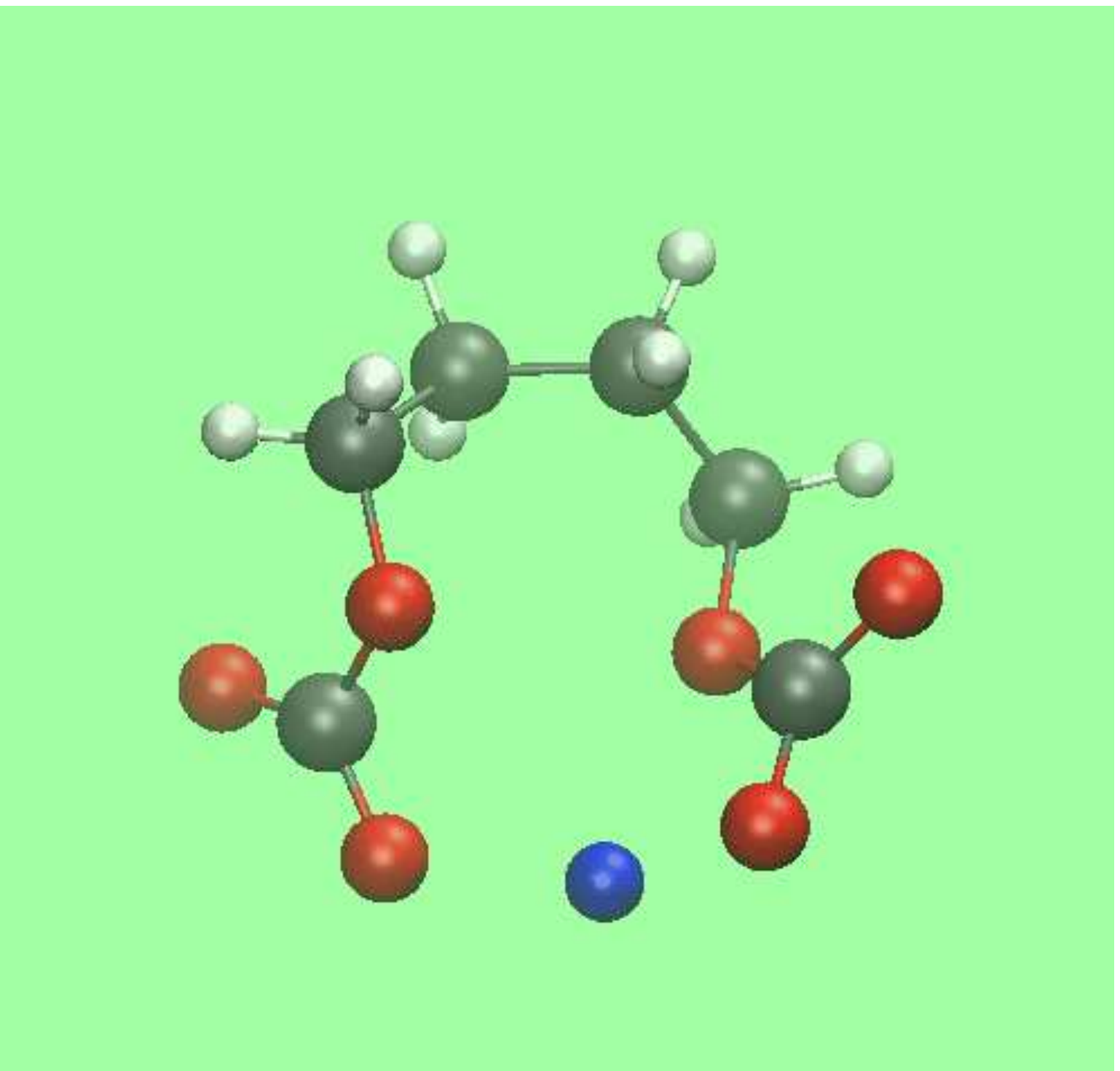}
		   (f) \includegraphics[width=1.40in]{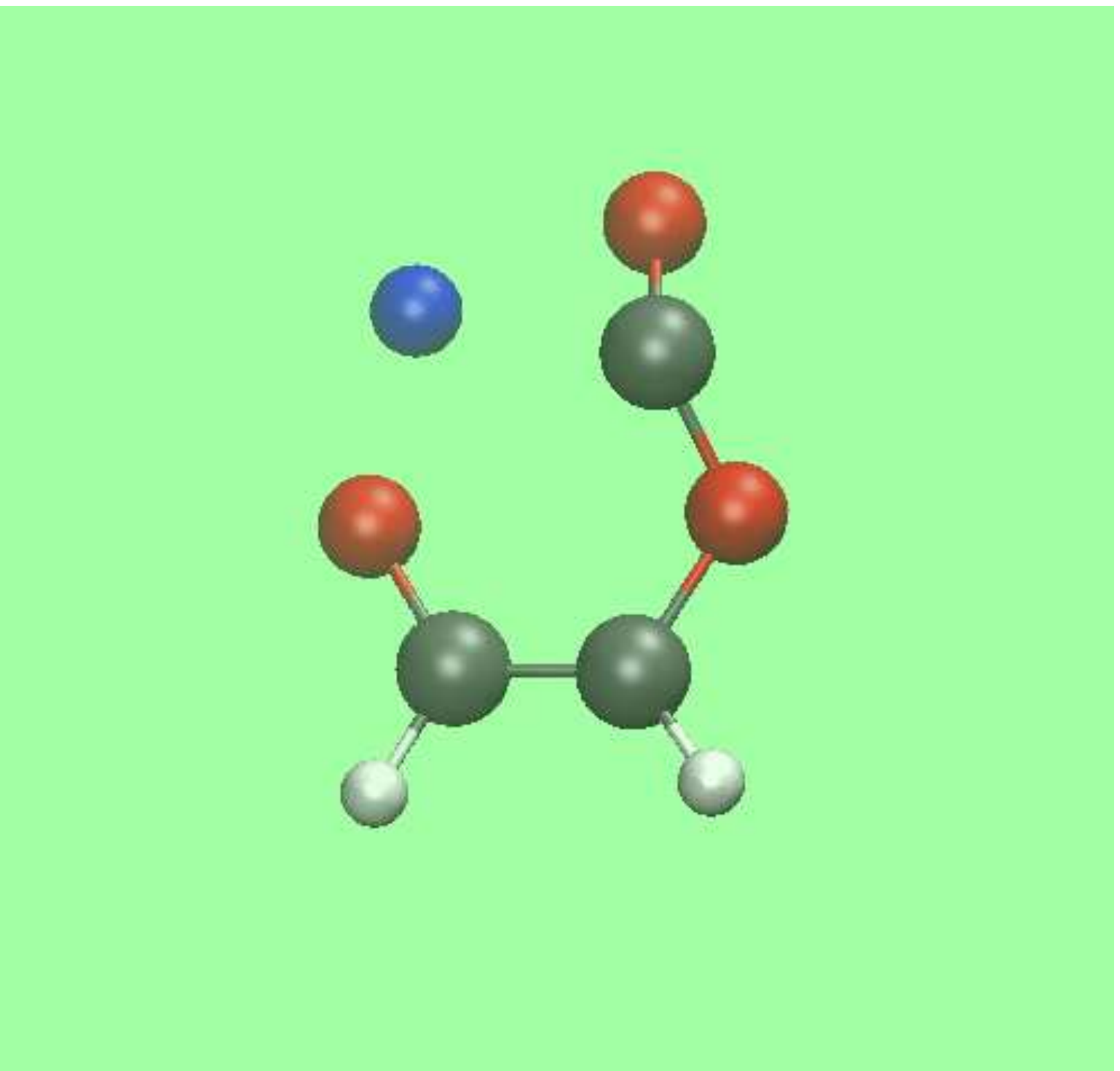} }}
\centerline{\hbox{ (g) \includegraphics[width=1.40in]{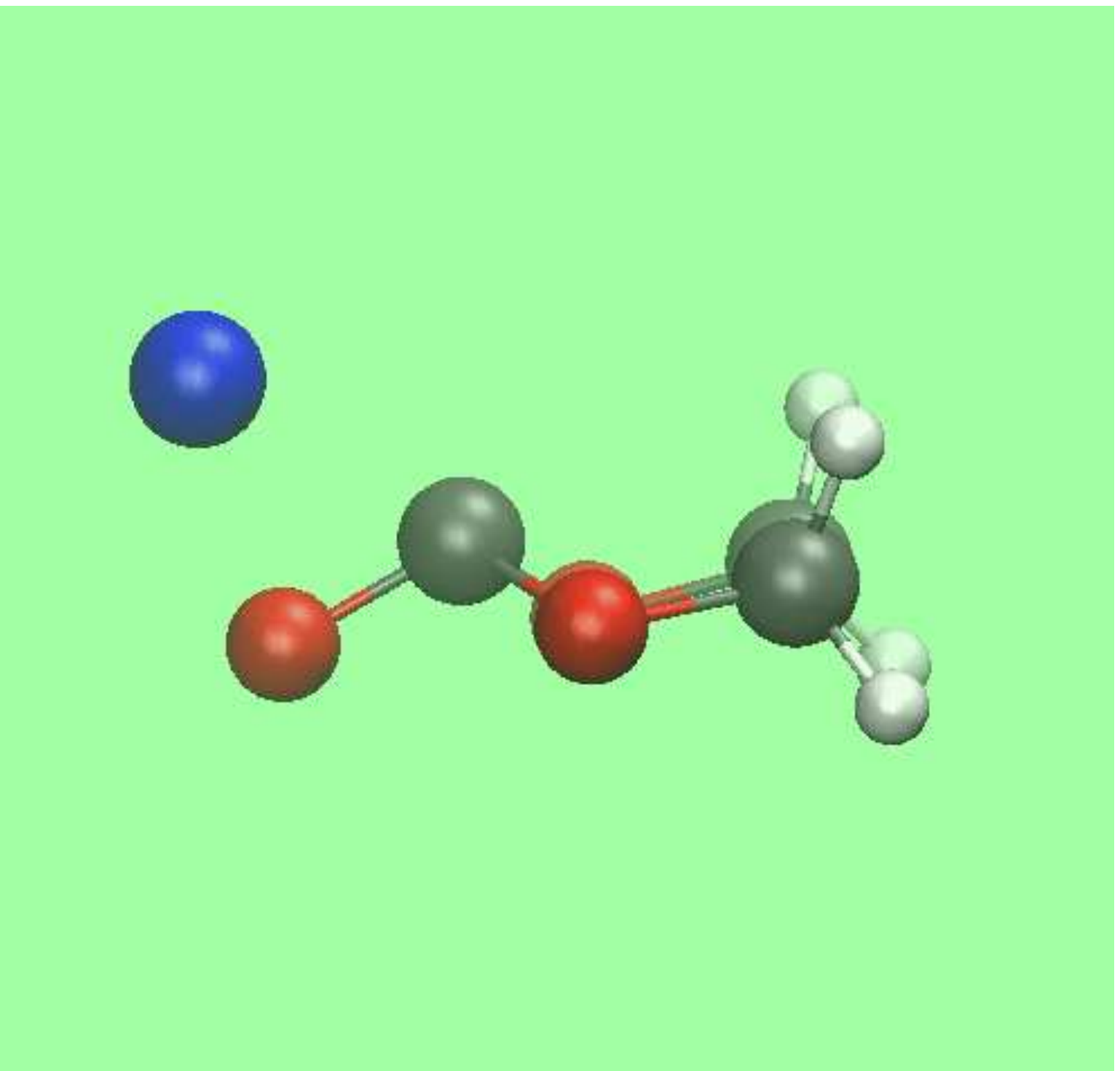}
		   (h) \includegraphics[width=1.40in]{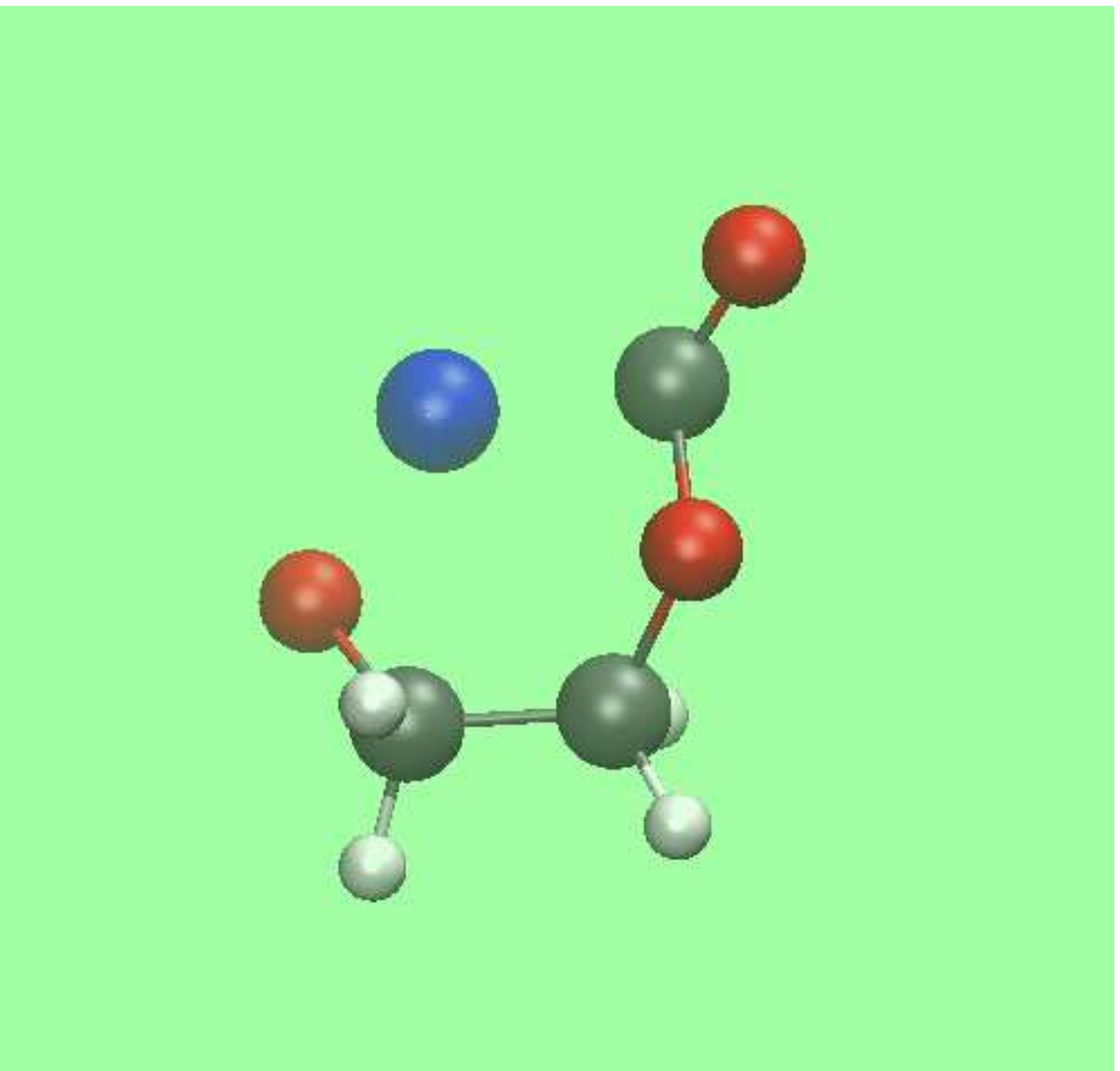}
		   (i) \includegraphics[width=1.40in]{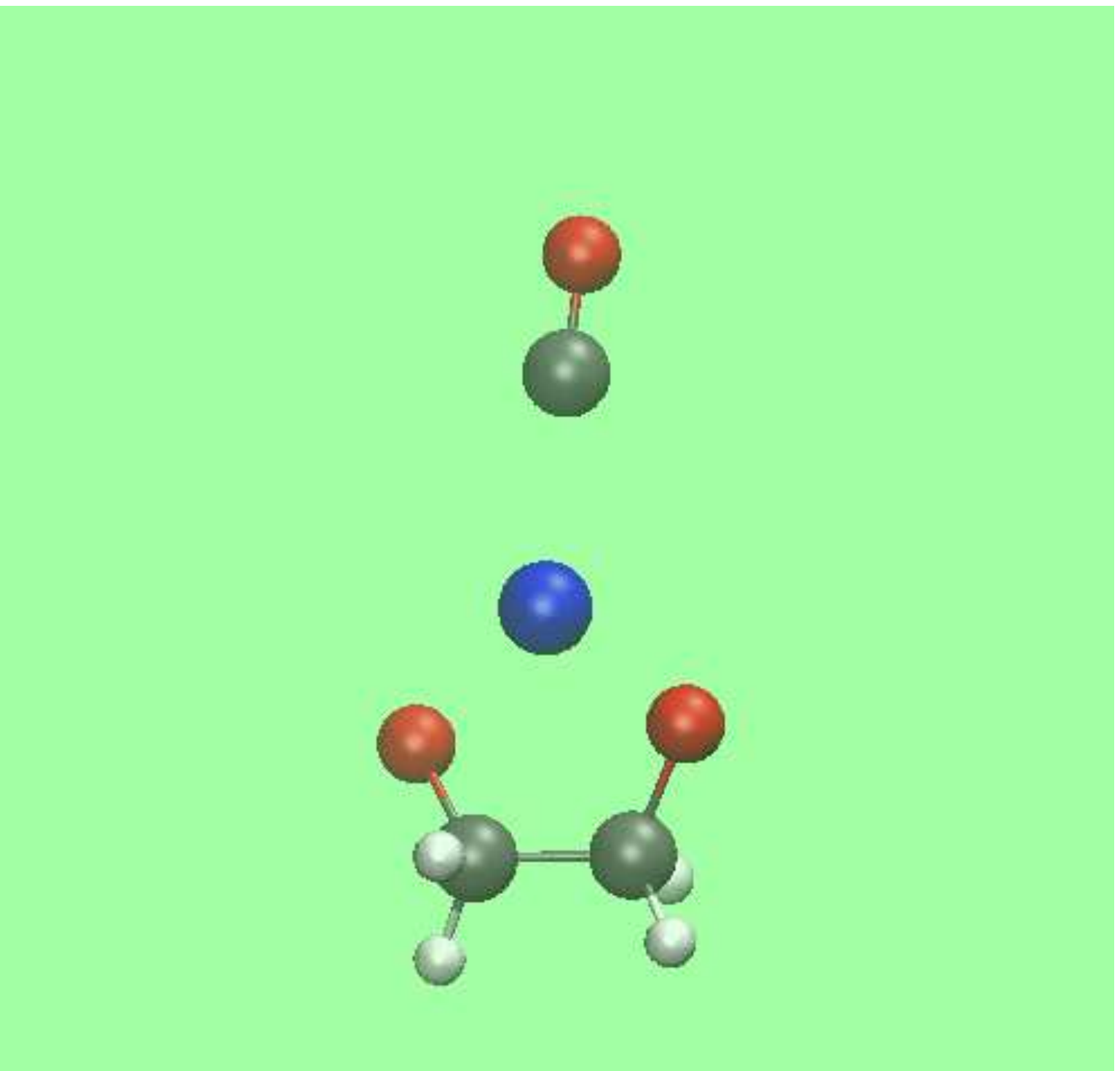} }}
\caption[]
{\label{fig1} \noindent
A pictorial review of proposed EC decompositon mechanisms.  (a) EC:Li$^+$
with atomic labels. (b) o-EC$^-$ radical anion, which arises from
exothermic, unimolecular cleavage of a C$_{\rm E}$-O$_{\rm E}$
bond in c-EC$^-$ (panel (c)).  The barrier is $\sim$0.5~eV. (d) It is widely
quoted that two o-EC$^-$ disproportionates into a C$_2$H$_4$ molecule plus
ethylene dicarbonate (EDC, shown coordinated to Li$^+$). (e)  However, a lay
organic chemist might predict that butylene dicarbonate BDC should be favored
because o-EC$^-$ radical recombination should exhibit a lower barrier.  (f)
Li$^+$:VC$^-$ with a broken C$_{\rm C}$-O$_{\rm E}$ bond.  Breaking this bond
in the gas phase is thermodynamically downhill and almost barrierless.
(g) Intact c-EC$^{2-}$. (h) EC$^{2-}$ with one broken C$_{\rm C}$-O$_{\rm E}$
bond, exothermic by 1.4~eV with a low (0.1~eV) barrier.  Unlike VC, two
$e^-$ are needed to break this bond in EC.  (i) EC$^{2-}$ with
two broken C$_{\rm C}$-O$_{\rm E}$ bonds, further exothermic by 0.27~eV with
a modest 0.22~eV barrier.  The panels represent optimized configuration
except for (d) and (e).  Grey, red, white, and blue spheres are
C, O, H, and Li atoms, respectively.
}
\end{figure}

Figure~\ref{fig1} and Eq.~\ref{old2e}-\ref{eg-2h} depict EC decomposition
mechanisms proposed in the literature.  (Oligomerization reactions, not as
widely studied, are omitted.\cite{tarascon,gewirth,yoshida,ogumi,bedrov})
\begin{eqnarray}
{\rm EC} + 2 e^- &\rightarrow& {\rm CO}_3^{2-} + {\rm C}_2{\rm H}_4 ; 
	\label{old2e} \\
{\rm EC} + e^- &\rightarrow& {\rm c-EC}^- ; \label{onee1} \\
{\rm c-EC}^-  &\rightarrow& {\rm o-EC}^- ; \label{onee2} \\
2 \hspace*{0.1in} {\rm o-EC}^-  &\rightarrow& {\rm EDC} + {\rm C}_2{\rm H}_4 ;
	\label{old1e} \\
2 \hspace*{0.1in} {\rm o-EC}^-  &\rightarrow& {\rm BDC} ; \label{bdc} \\
{\rm c-EC}^- + e^-&\rightarrow& {\rm OC}_2{\rm H}_2{\rm O}^{2-}+ {\rm CO} .
			\label{eg-2h} 
\end{eqnarray}
When electron transfer is fast, it has been suggested that two electrons
transfer to the same EC to yield CO$_3^{2-}$ and C$_2$H$_4$
(Eq.~\ref{old2e}).\cite{review}  This reaction route appears consistent
with the evolution of C$_2$H$_4$ (among other gases) observed in gas
chromatography.\cite{onuki,ota1,ota2,shin} Li$_2$CO$_3$ has also been
reported as part of the SEI.  However, at least one experimental group observes
rather small amount of Li$_2$CO$_3$ when air is avoided during sample
transfer.  One rationale proposed is that lithium carbonate could arise from
decomposition of other SEI components during ex-situ analysis.\cite{edstrom}

A one-electron mechanism (Eqs.~\ref{onee1}-\ref{old1e}) is often assumed to
dominate when the electron tunneling from anode to electrolyte is
slow.\cite{review,intro1}  It involves o-EC$^-$ radical anions formation
from intramolecular ring-opening C$_{\rm E}$-O$_{\rm E}$ cleavage of c-EC$^-$
(Fig.~\ref{fig1}b,~\ref{fig1}c; Eq.~\ref{onee2}).  It has been widely
quoted that two o-EC$^-$ disporportionate to yield ethylene dicarbonate
(EDC, Fig.~\ref{fig1}d) and C$_2$H$_4$ (Eq.~\ref{old1e}) afterwards.  EDC
has been identified as the main component of SEI on graphitic anodes in
FTIR measurements,\cite{edc} and species with unpaired spins have been
observed under half-cell conditions.\cite{endo} These observations have
been cited to support Eq.~\ref{old1e}.

In terms of theory, Balbuena {\it et al.} has applied hybrid DFT functionals
to examine EC decomposition after injection of one excess electron
(Eq.~\ref{onee1}).\cite{bal01}  The c-EC$^-$$\rightarrow$o-EC$^-$
reaction (Eq.~\ref{onee2}), breaking a C$_{\rm E}$-O$_{\rm E}$ bond, is
predicted to exhibit a 0.49~eV barrier with an approximate dielectric
solvation model.  This implies a millisecond reaction rate at room temperature.
Quantum chemistry techniques like MP2 and CCSD(T) predict a slightly higher
barrier.\cite{han2004}  The DFT/PBE functional predicts a lower but
still significant, 0.4~eV gas phase barrier.\cite{pccp} However,  AIMD/PBE
simulations of liquid EC yield an anomalously fast C$_{\rm E}$-O$_{\rm E}$
cleavage rate for EC$^-$ {\it not} coordinated to Li$^+$ only, partly due to
physical many-solvent effects\cite{bal11} and partly to unphysical,
multi-electron self-interaction errors.\cite{wtyang1} AIMD using more accurate
hybrid functionals have not shown such fast EC$^-$ decomposition.\cite{bal11}
In the absence of costly hybrid functional AIMD predictions of reaction rate
in the liquid state, we tentatively assume that Eq.~\ref{onee2} exhibits a
reaction rate of $\sim$10$^3$/s for the purpose of this review, but note
that the presence of electrode surfaces can significantly increase this
rate (Fig.~3 \&~6, Ref.~\onlinecite{ald}).

Eq.~\ref{old1e}, the widely quoted second half of the one-electron mechanism,
has received less theoretical study.  Balbuena {\it et al.} has found
that it is less exothermic than Eq.~\ref{bdc}.  Its activation energy cannot
be estimated using cluster-based DFT methods which struggle with two
spin-antiparallel o-EC$^-$ radicals.  Using the ReaxFF force field, Bedrov
{\it et al.} have recently estimated a 0.61~eV (14~kcal/mol) gas phase barrier
for Eq.~\ref{old1e} compared with a much lower 0.15~eV (3.5~kcal/mol) barrier
for butylene dicarbonate (BDC) formation (Eq.~\ref{bdc}, 
Fig.~\ref{fig1}e).\cite{bedrov}  ReaxFF is fitted to electronic structure
data and ignores electron spin.  Despite its approximate nature, it agrees
with the DFT prediction\cite{bal01} that BDC is more stable than EDC plus
C$_2$H$_4$.  More significantly, EDC formation is now predicted to be both
kinetically and thermodynamically unfavorable.  This ReaxFF prediction
dovetails with the intuitive organic chemistry viewpoint that recombination of
sterically unhindered radicals is generally barrierless and kinetically
favored over bond-breaking reactions like Eq.~\ref{old1e}.  While Bedrov
{\it et al.} report one instance of EDC formation in liquid EC, that
calculation starts with an extremely high EC$^-$ radical concentration
which is more consistent with the two-electron regime (see below),
not the one-electron regime thought to be conducive to EDC formation.  

It is therefore fair to state the following: theoretical studies have
consistently shown that Eqs.~\ref{onee1}~and~\ref{onee2} constitute a pathway
for BDC (Eq.~\ref{bdc}), {\it not} EDC (Eq.~\ref{old1e}), formation.  This has
practical consequences.  SEI films continue to grow during power cycling and/or
storage over a battery's lifetime.  The widely accepted Eq.~\ref{old1e}
suggests that a C$_2$H$_4$ gas molecule is released per two electron/two Li$^+$
consumed in continued SEI growth.  If 0.1~mole of Li$^+$ is consumed inside a
pouch cell over its lifetime, close to a liter of C$_2$H$_4$ will be released
according to Eq.~\ref{old1e}.  This seems untenable.  And if Eq.~\ref{old1e}
does not occur, we know of no intrinsic one-electron mechanism consistent with
the formation of C$_2$H$_4$ or EDC observed during SEI formation.

As will be discussed, a two-electron route actually leads to low-barrier
EDC formation under some conditions.  But first we take a detour to vinylene
carbonate (VC) for soon-to-be self-evident reasons.  VC is a popular additive
to battery electrolytes.  Upon reduction, its C=C motif yields polymeric
products that improve the SEI.  Balbuena and coworkers have predicted that the
C$_{\rm E}$-O$_{\rm E}$ bond breaking barrier of VC$^-$ is higher than that
of c-EC$^-$.\cite{bal02}  Using CCSD(T) methods, Han and Lee also predict a
1.07~eV barrier.\cite{han2004}  However, in the gas phase, breaking another C-O
bond, C$_{\rm C}$-O$_{\rm E}$ (Fig.~\ref{fig1}f), is downhill in energy and is
practically barrierless when zero-point energy is accounted for.  This appears
the first report indicating the fragility of the C$_{\rm C}$-O$_{\rm E}$
bond in reduced cyclic carbonates.  (Note to experimentalist readers: large
basis set ``CCSD(T)'' calculations are more systematic and trustworthy than
DFT with approximate functionals, although extremelly costly.  When correctly
applied to small molecules made of second-row element, this author consider
CCSD(T) infallible.) When an approximate dielectric solvation model is added,
our MP2 calculations predict that C$_{\rm C}$-O$_{\rm E}$ is thermoneutral
($\Delta G$=0.004~eV).  DFT calculations have suggested a slightly more
positive $\Delta G$=0.16~eV (Ref.~\onlinecite{book}, Ch.~5).  Regardless of
the precise exothermicity of this intermediate, breaking the remaining
C$_{\rm C}$-O$_{\rm E}$ bond in VC$^-$ to yield CO is very exothermic (1.73~eV
via MP2 predictoins).  This is consistent with the substantial CO gas released
in experiments when VC is the sole organic solvent.\cite{ota1,ota2} (CO$_2$
is also released via proposed VC decarboxylation; other reactions do not
release any gas.\cite{bal02a})

This brings us back to EC.  Using CCSD(T), the C$_{\rm C}$-O$_{\rm E}$ bond
cleavage in the EC$^-$:Li$^+$ complex is predicted to exhibit a small, 0.09~eV
barrier.\cite{han2004}  Unlike VC, the reaction is thermodynamically uphill
even in the gas phase.  However, our previous work has shown that a second
excess $e^-$, the presence of an electrode at low voltage, and/or EC$^-$ 
coordination to a material surface that favors bond-breaking can enable this
fast decomposition route.\cite{pccp,bal11,ald}  Here we apply the MP2 method,
which gives EC$^-$ energetics similar to CCSD(T),\cite{han2004} to consolidate
Eq.~\ref{onee1} and~\ref{eg-2h}.  A dielectric continuum approximation is
used.  Fig.~\ref{fig1}g and~\ref{fig1}h depict a c-EC$^{2-}$ and a EC$^{2-}$
with a broken C$_{\rm C}$-O$_{\rm E}$ bond, respectively.  The reaction is
exothermic by 1.4~eV with a barrier of only 0.10~eV.  Breaking the second
C$_{\rm C}$-O$_{\rm E}$ bond yields a modest 0.22~eV barrier and 0.27~eV
exothermicity, and leads to release of CO (Fig.~\ref{fig1}i.) The low
C$_{\rm C}$-O$_{\rm E}$ cleavage barriers are qualitatively consistent with
CO formation seen in AIMD/PBE simulations of explicit liquid
electrolyte/electrode interfaces.\cite{pccp}  A substantial amount of CO
gas has been reported in experiments\cite{onuki,ota1,ota2,yoshida,shin}
although some measurements report less of it.\cite{novak1998,intro1}

In the literature, the other 2-$e^-$ reaction, breaking the
C$_{\rm E}$-O$_{\rm E}$ bonds in EC$^{2-}$:Li$^+$ to yield CO$_3^{2-}$ and
C$_2$H$_4$ (Eq.~\ref{old2e}) has received far more
attention.\cite{bal01,vollmer}  Our MP2 calculations predict a 0.40~eV
bond-breaking barrier for Eq.~\ref{old2e}, much higher than that for cleaving
C$_{\rm C}$-O$_{\rm E}$ bonds to form CO.  Although less thermodynamically
favorable than Eq.~\ref{old2e}, the CO-releasing reaction (Eq.~\ref{eg-2h}
must dominate during kinetically controlled, irreversible SEI growth
conditions unless other factors (e.g., electrode surfaces) are involved.  Note
also that the breaking of an analogous C$_{\rm C}$-O bond in DMC, a linear
co-solvent, has been widely accepted because CH$_3$O$^-$ species have been
detected in the SEI.\cite{review}  Breaking chemically similar C$_{\rm C}$-O
bonds in cyclic carbonates like EC should clearly be viable under the same
experimental conditions.  

\begin{table}\centering
\begin{tabular}{||l|l|l||} \hline
oxidized & reduced & $\Phi$ (V) \\ \hline
EC & c-EC$^-$ & $+$0.02 \\ 
c-EC$^-$ & c-EC$^{2-}$ & +0.66 \\ \hline 
EC:Li$^+$ & c-EC$^-$:Li$^+$ & +0.53 \\ 
c-EC$^-$:Li$^+$ & c-EC$^{2-}$:Li$^+$ & +1.16 \\ \hline 
(EC)$_3$Li$^+$ & (EC)$_2$c-EC$^-$Li$^+$ & +0.55 \\ %
(EC)$_2$c-EC$^-$Li$^+$ & (EC)$_2$c-EC$^{2-}$Li$^+$ & +1.06 \\ \hline
\end{tabular}
\caption[]
{\label{table1} \noindent
Reduction potentials ($\Phi$) vs Li$^+$/Li(s) (i.e., subtract 1.37~eV), in volt.
Finite temperature corrections are included and $\epsilon$=40.
}
\end{table}

\section{Reduction potentials favor two-electron attack}
\label{reduction}

Next we stress that two electron attacks on EC are favorable from a
electron transfer perspective.\cite{ald}  We show that excess electrons
attack EC in descending order of preference:
\begin{equation}
{\rm o}{\rm -}{\rm EC}^- > {\rm c}{\rm -}{\rm EC}^- > {\rm EC} . \label{order}
\end{equation}
The geometries of (EC)Li$^+$, (c-EC$^-$)Li$^+$, and (c-EC$^{2-}$)Li$^+$
clusters, solvated with $\epsilon$=40, are optimized (not shown).  From the
predicted free energies, the reduction potentials ($\Phi$) of EC and
c-EC$^-$ are 0.53 and 1.16~V relative Li$^+$/Li(s), respectively
(Table~\ref{table1}).  Thus c-EC$^-$ accepts an electron much more readily.
Similar relative $\Phi$ are predicted for the (EC)$_3$Li$^+$ cluster and even
for an isolated EC$^{n-}$ molecule solvated in the dielectric continuum,
although the absolute $\Phi$ are much less favorable in that last case
(Table~\ref{table1}).  Note that MP2 predicts lower $\Phi$ values than DFT
methods.\cite{vollmer}  Some researchers have been suggested that the reduction
onset voltage should actually reflect EC$\rightarrow$o-EC$^-$.\cite{vollmer}
However, in bulk liquid, bond-breaking to form o-EC$^-$ seems to exhibits a
substantial barrier.\cite{bal01}  EC$^-$ ring-opening must be substantially
accelerated, e.g., via the presence of an electrode, to make o-EC$^-$
relevant to voltametry measurements.

\begin{figure}
\centerline{\hbox{ (a) \includegraphics[width=2.00in]{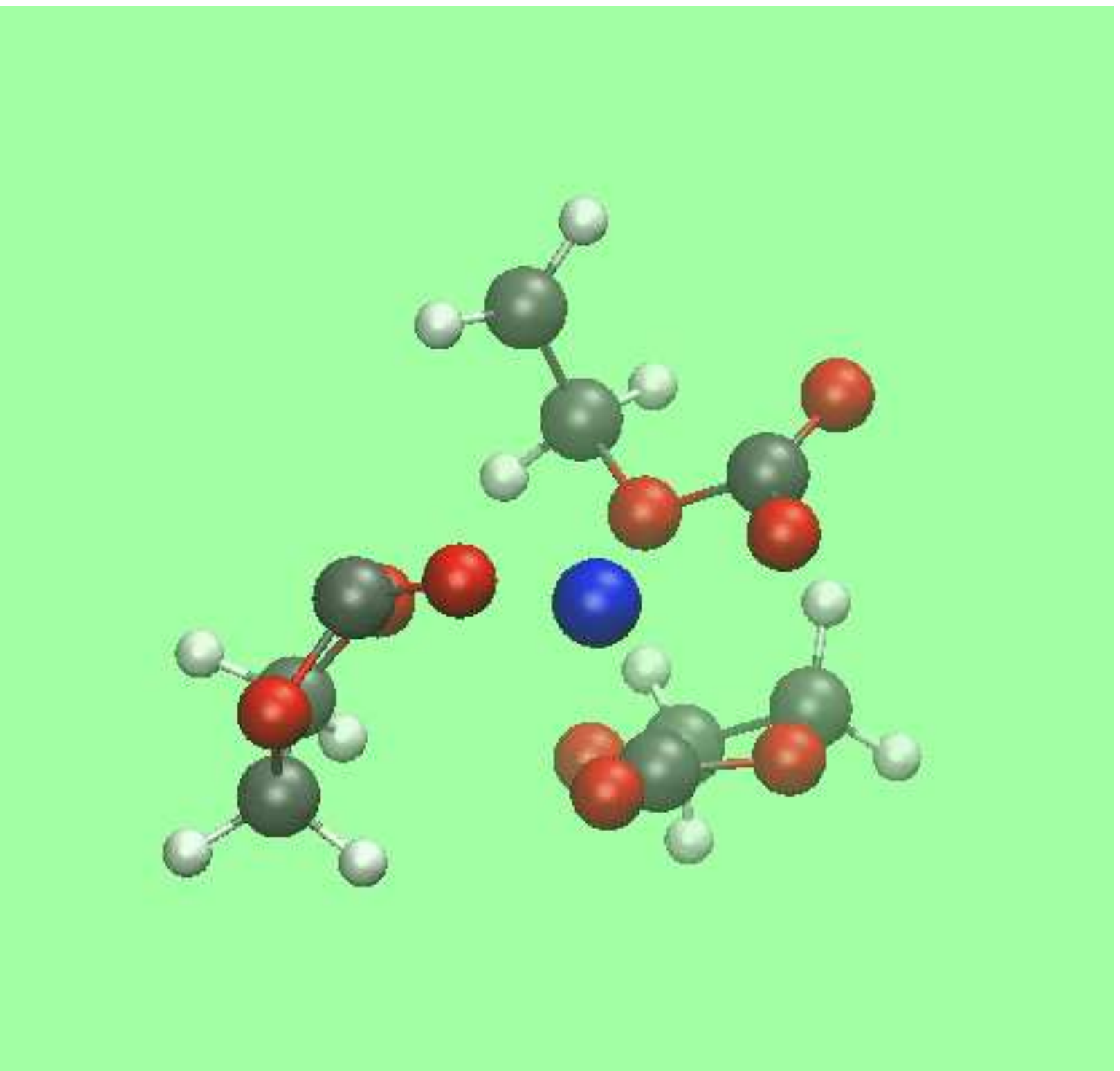}
                   (b) \includegraphics[width=2.00in]{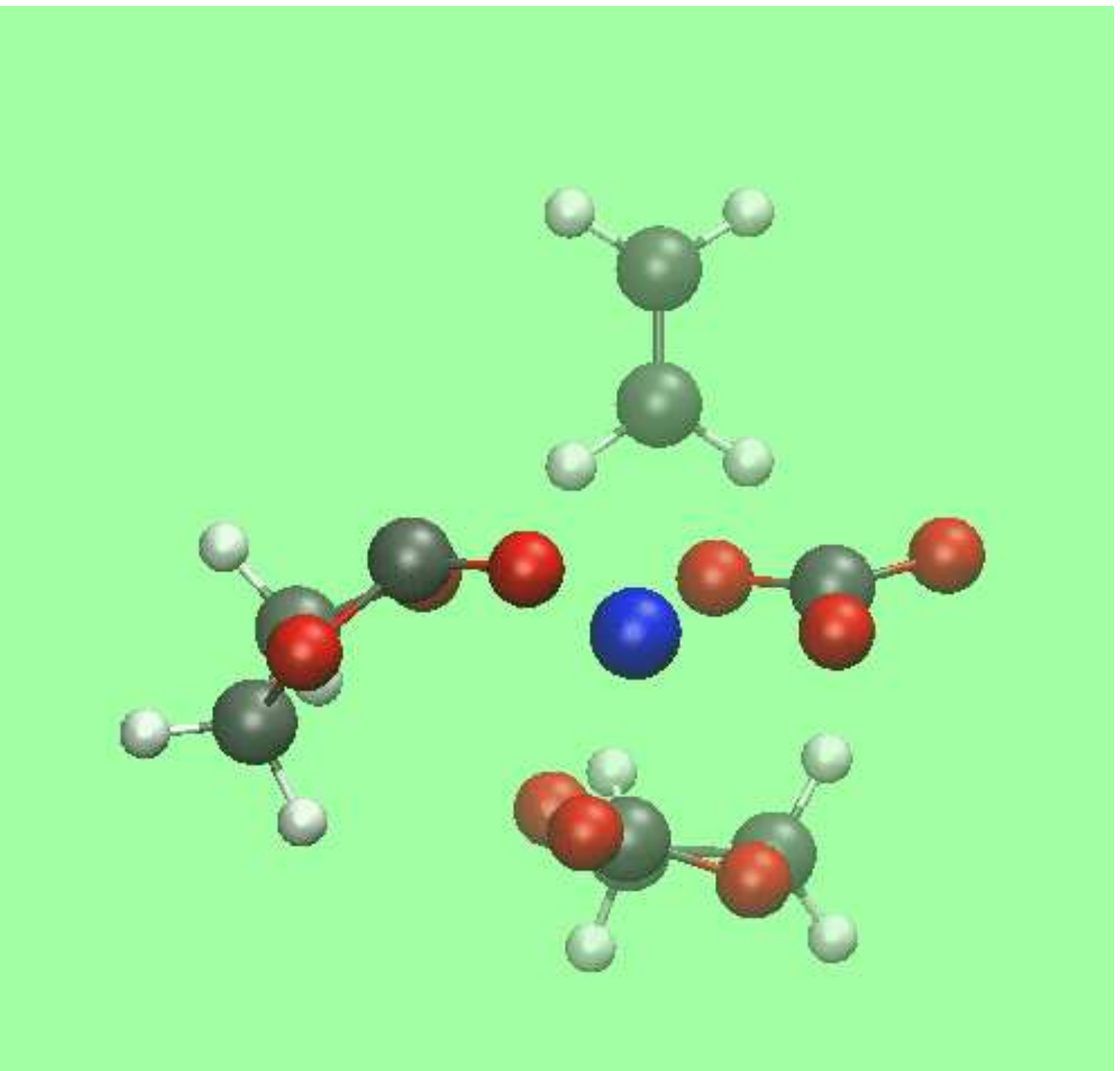} }}
\caption[]
{\label{fig2} \noindent
(a) Starting with optimized spin-triplet (EC)(c-EC$^-$)(o-EC$^-$)Li$^+$ and 
adding an extra electron lead to (EC)(c-EC$^-$)(o-EC$^{2-}$)Li$^+$.  (b) The
o-EC$^{2-}$ fragment then decomposes into C$_2$H$_4$ and CO$_3^{2-}$.
}
\end{figure}

These estimates are pertinent to $k_e'$$>$$k_3$$\sim$10$^3$/s.  At lower
$e^-$ tunneling rate, c-EC$^-$ has time to decompose into o-EC$^-$ before
accepting another $e^-$.  To compare o-EC$^-$ and c-EC$^-$ reduction
qualitatively, we optimize a spin-triplet (c-EC$^-$)(o-EC$^-$)(EC)Li$^+$
cluster with $\epsilon$=40 (Fig.~\ref{fig2}a).  This spin configuration forces
the two excess electrons to reside on different EC molecules.  Next, an extra
$e^-$ is added.  Mulliken analysis shows that two excess $e^-$ reside on
o-EC$^-$, a single $e^-$ on the bent c-EC, and none on the charge neutral EC.
This shows that o-EC$^-$ accepts an $e^-$ more readliy than c-EC$^-$.  Upon
geometry optimization, o-EC$^{2-}$ spontaneously breaks up into CO$_3^{2-}$
and C$_2$H$_4$ (Fig.~\ref{fig2}b, Eq.~\ref{old2e}) without an apparent
kinetic barrier.  Henceforth we focus on the $e^-$ transfer to c-EC$^{-}$.
$e^-$ addition to o-EC$^{-}$ is faster, and the c-EC$^-$ reduction rate
used will be a lower-bound for o-EC$^-$ reduction.

\begin{figure}
\centerline{\hbox{ (a) \includegraphics[width=3.00in]{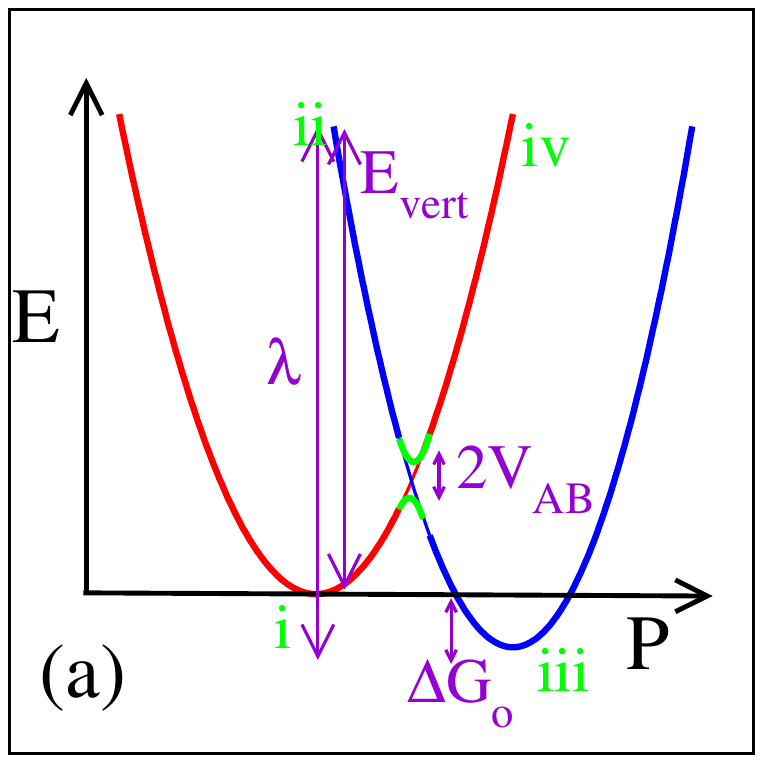}
                   (b) \includegraphics[width=2.00in]{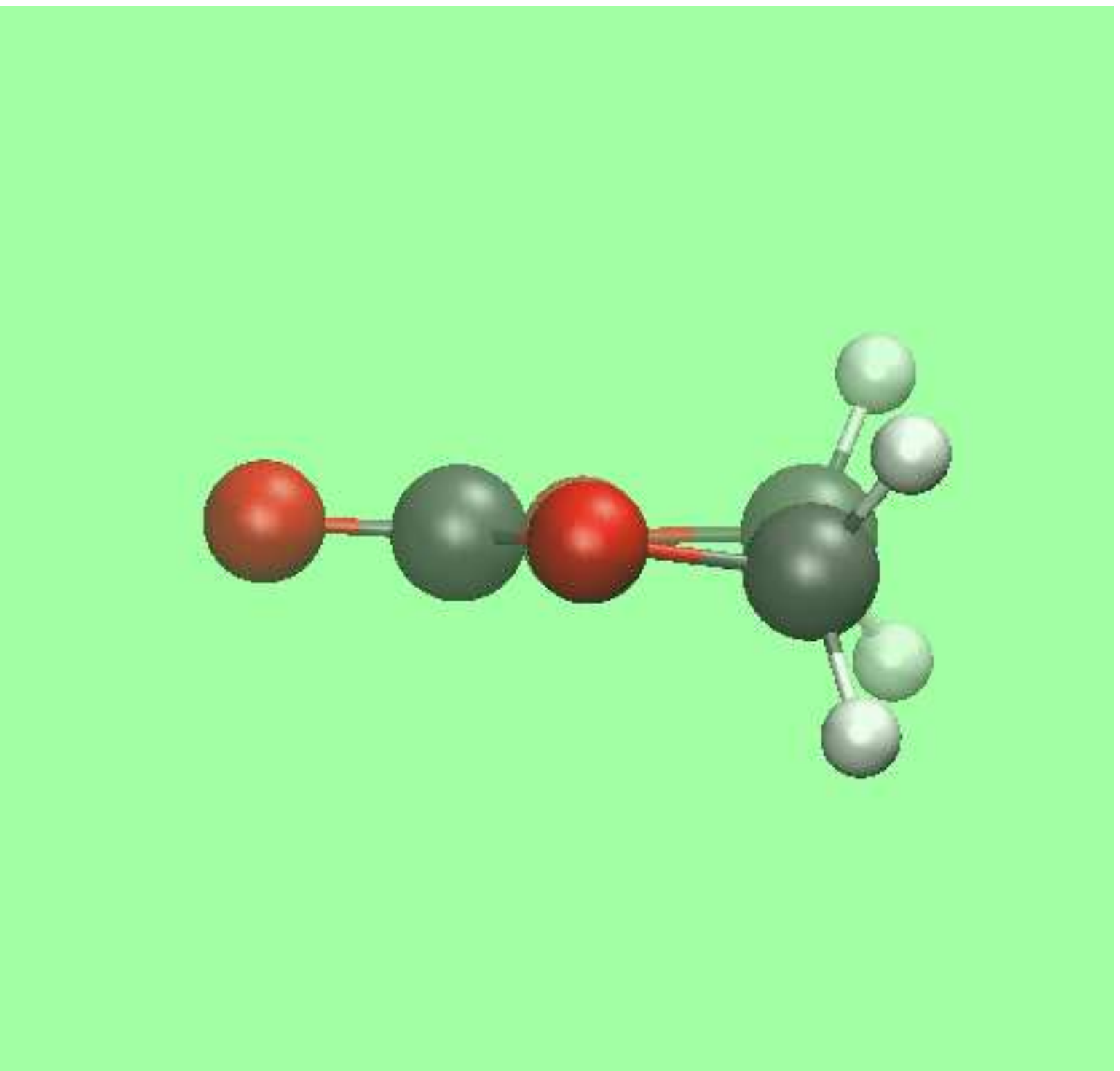} }}
\centerline{\hbox{ (c) \includegraphics[width=2.00in]{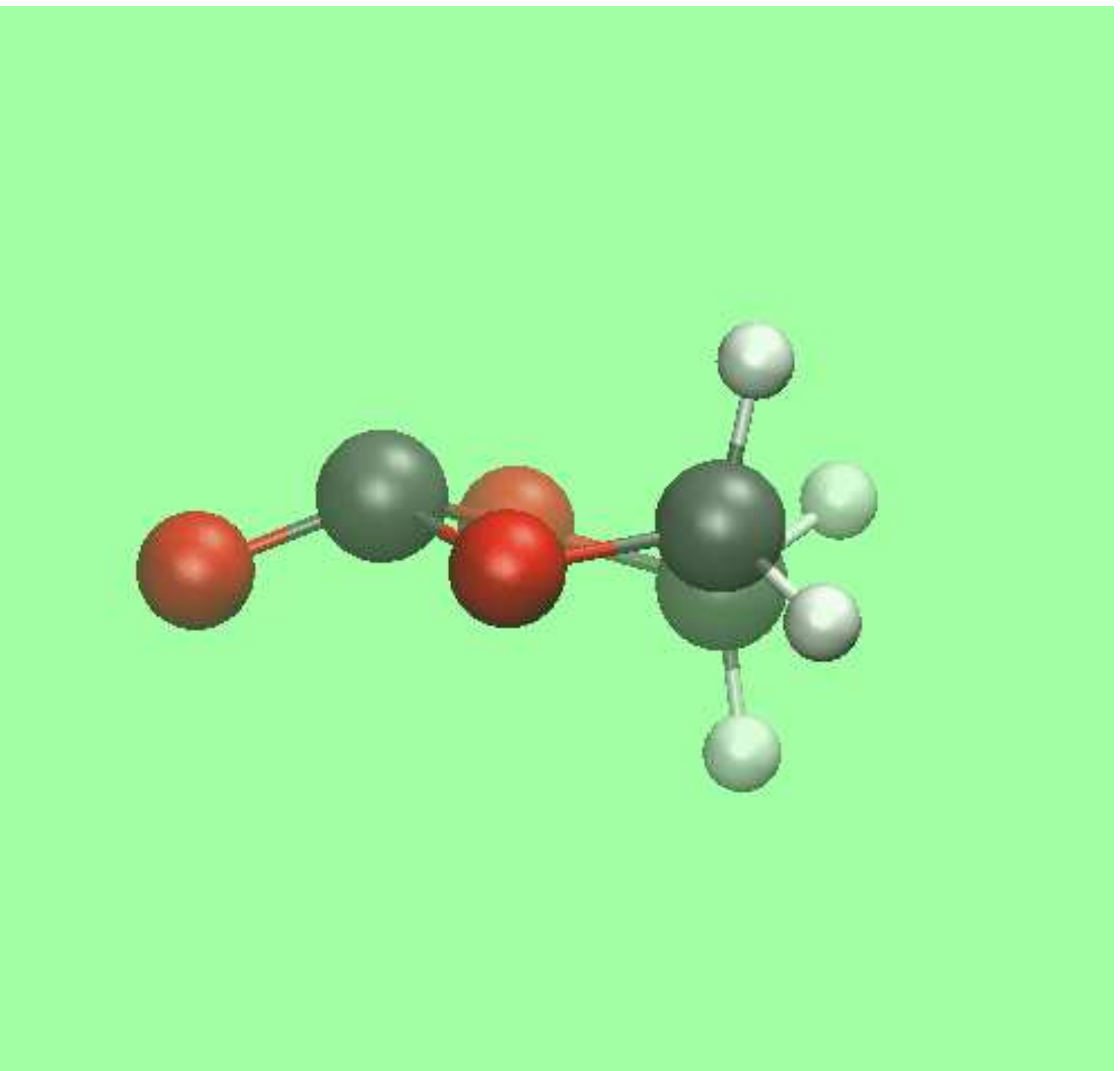}
                   (d) \includegraphics[width=2.00in]{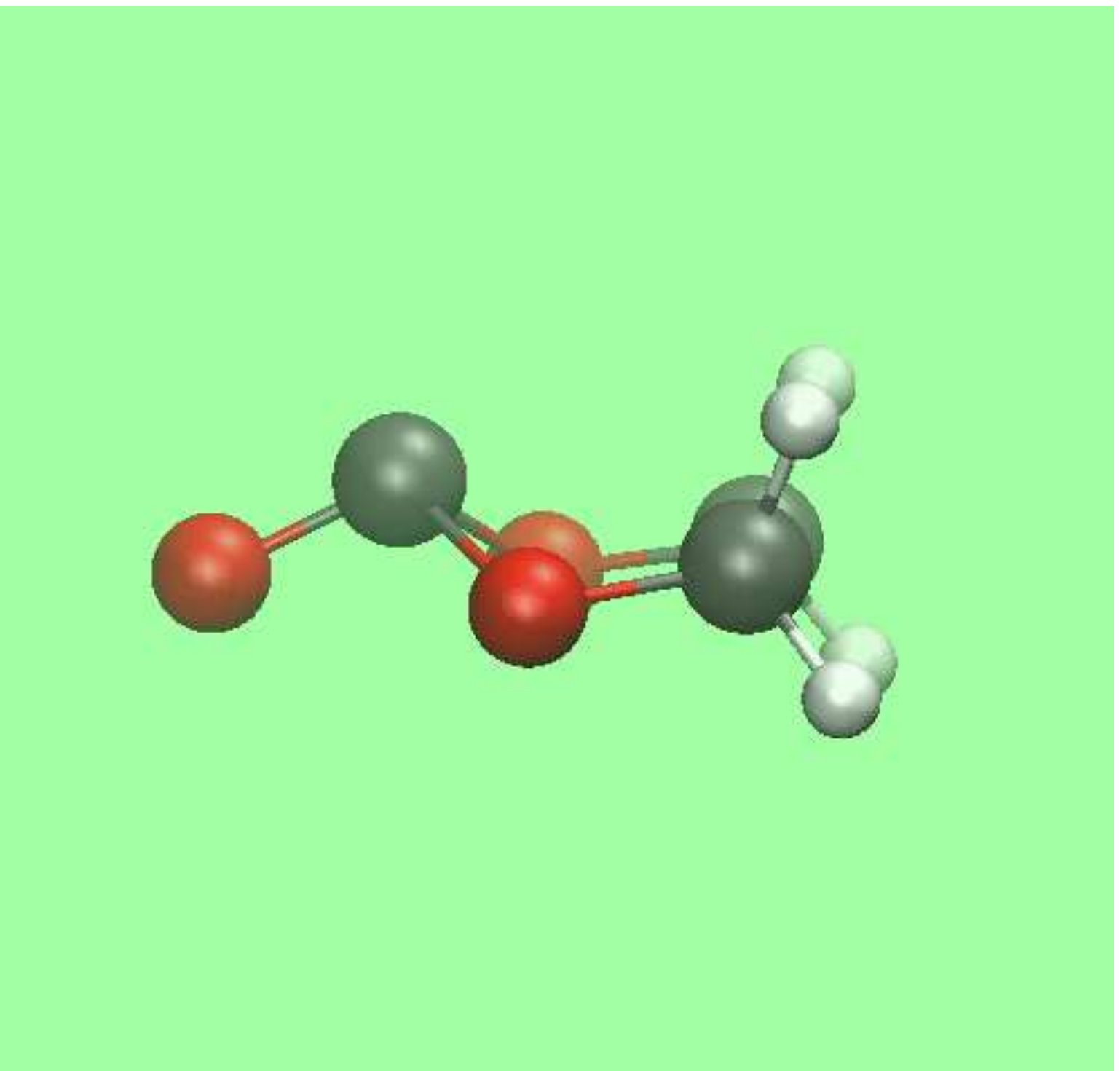} }}
\caption[]
{\label{fig3} \noindent
(a) Standard Marcus theory parabolic construction; (b) EC, which exhibits
an almost flat ring geometry; (c) c-EC$^-$, bent geometry; (d) c-EC$^{2-}$,
a higher degree of bending than c-EC$^-$.  All structures are optimized with
$\epsilon$=40.  For EC reduction, (i,ii) and (iii,iv) have EC and c-EC$^-$
atomic structures respectivevly; for c-EC$^-$ reduction, (i,ii) and (iii,iv)
have c-EC$^-$ and c-EC$^{2-}$ atomic structures.  The red and blue curves
indicate the oxidized and reduced potential energy surfaces. (ii) and (iv)
require the use of the high frequency dielectric constant.
}
\end{figure}
\section{Reorganization energies}
\label{reorg}

The non-adiabatic rate for electron transfer into
EC$^{n-}$ molecules is governed by\cite{marcus}
\begin{equation}
k \sim V_o^2 \exp \{ - \beta (\lambda + \Delta G^o)^2/4\lambda \},\label{marcus}
\end{equation}
where $V_o$ is the tunneling matrix element, $\lambda$ is the reorganization
energy, and $\beta$ is $1/(k_{\rm B}T)$.  $\Delta G^o$ is 
$-q_e\Phi$ added to the applied underpotential.  This section argues that
reorganization energies $\lambda$ also favor doubly charged c-EC$^{2-}$.
Due to its more speculative and technical nature, the casual reader is
encouraged to skip over to Sec.~\ref{crossover}.

Reorganization energies ($\lambda$) are estimated via a Marcus theory
construction (Fig.~\ref{fig3}a).  First we consider a heuristic model: a
single EC molecule solvated by a $\epsilon$=40 dielectric continuum
(Fig.~\ref{fig2}).  This ignores the fact that only fast electronic
polarization changes in outershell liquid EC molecules can response to
electron injection.  Thus the $\lambda$ reported in Table~\ref{table2} can
be construed as entirely due to intramolecular contributions.  Within Marcus
theory, $\lambda$ for the forward and backward electron transfer processes
should match.  In our simple estimates, they can disagree by up to~33\%
for both EC$\rightarrow$c-EC$^-$ and
c-EC$^-$$\rightarrow$c-EC$^{2-}$.\cite{marcus}  In this case we simply
take the mean of the forward and backward $\lambda$.  Then $\lambda$
averages to 1.78 and 0.82~eV for adding the first and second $e^-$ to an
intact EC.  Similar values (1.80 and 0.62~eV) are obtained if the high
frequency $\epsilon$=2.62 is adopted for all configurations, or if Li$^+$ are
present in the calculations (Table~\ref{table2}, but see caveats below).

The lower intramolecular $\lambda$ for c-EC$^-$ reduction has a simple
structural rationale.  The C-O framework of neutral EC is almost flat.  The
$sp^2$-hybridized C$_{\rm C}$ carbon atom needs to switch to $sp^3$, with the
C$_{\rm C}$-O$_{\rm C}$ bond bent out of plane, to accommodate an excess
$e^-$, entailing a large $\lambda$.  c-EC$^-$ is already bent
(although not as much as c-EC$^{2-}$) and $sp^3$ hybridized (Fig.~\ref{fig2});
therefore the c-EC$^-$ to c-EC$^{2-}$ reorganization should be much smaller.

As mentioned above, additional outershell contributions to $\lambda$ exist.
Dielectric continuum-based $\lambda$ calculations which attempt to account for
such frequency-dependent effects yield predictions which differ from
experiments by a substantial fraction of an electron volt.\cite{voorhis}
Nevertheless, we apply such a rough estimate for the full $\lambda$'s.
We somewhat arbitarily choose charge-neutral clusters with one EC and
select the following relaxations within an electronic surface following
a vertical excitation
(Fig.~\ref{fig3}a):
\begin{eqnarray}
( {\rm c-EC}^{2-}{\rm Li}^+ )^0 &\rightarrow& ( {\rm c-EC}^{-}{\rm Li}^+ )^0 \\
\label{lambda1}
( {\rm EC}{\rm Li}^+ )^0  &\rightarrow&  ( {\rm c-EC}^{-}{\rm Li}^+ )^0 ,
\label{lambda2}
\end{eqnarray}
where the left and right sides are computed using $\epsilon$=2.62 and
$\epsilon$=40.0, respectively.  This yields overall $\lambda$=1.61 and 2.05~eV
for reduction of c-EC$^-$ and EC, respectively.  Despite the potentially
large uncertainties associated with these estimates, $\lambda$ for reduction of
c-EC$^-$ is again predicted to be favored.  Future AIMD potential-of-mean-force
simulations should be used to yield improved $\lambda$ values.
\begin{table}\centering
\begin{tabular}{||l|l|l|l||} \hline
ground & excited & $\lambda_{\epsilon=2.62}$ & $\lambda_{\epsilon=40}$ \\ 
						\hline
EC & [c-EC$^-$]$^+$ & 1.863 & 1.916 \\
c-EC$^-$ & [c-EC]$^-$ & 1.731 & 1.639 \\ \hline
c-EC$^-$ & [c-EC$^{2-}$]$^+$ & 0.427 & 0.657 \\
c-EC$^{2-}$ & [c-EC$^-$]$^-$ & 0.802 & 0.980 \\ \hline
EC:Li$^+$ & [c-EC$^-$]$^+$:Li$^+$ & 2.018 & 2.000 \\
c-EC$^-$:Li$^+$ & [c-EC]$^-$:Li$^+$ & 1.383 & 1.543 \\ \hline
c-EC$^-$:Li$^+$ & [c-EC$^{2-}$]$^+$:Li$^+$ & 0.723 & 0.987 \\
c-EC$^{2-}$:Li$^+$ & [c-EC$^-$]$^-$:Li$^+$ & 0.955 & 0.960 \\ \hline
\end{tabular}
\caption[]
{\label{table2} \noindent
Intramolecular reorganization energies ($\lambda$) with $\epsilon$ kept
constant.  ZPE is not included.
}
\end{table}

(EC)$_3$Li$^+$ is not used to calculate $\lambda$ because adding an $e^-$ to
this cluster at $\epsilon$=2.62 erroneously leads to the excess $e^-$ residing
on the Li$^+$ ion.  In more realistic, periodic boundary condition AIMD
simulations of liquid EC, instantaneously adding an $e^-$ results in a
delocalized $e^-$ in the conduction band,\cite{pccp,bal11} never on the
Li$^+$ ion.  (Cluster models do not have a liquid state conduction band.)
Adding $e^-$ to particular EC molecules requires a constrained-DFT/AIMD
approach.\cite{ald,voorhis}  Henceforth we will use $\lambda$ and $\Phi$ from
c-EC$^{n-}$:Li$^+$ clusters.  Another caveat is that, 
for $\lambda$ calculations only, we have used a c-EC$^-$:Li$^+$ cluster
with the Li lying on the bisector of the molecule.  It is relaxed from
Fig.~\ref{fig1}g by removing an $e^-$.  This configuration is 0.081~eV
higher in energy than that of Fig.~\ref{fig1}c.  (The energy difference is
much smaller -- within 0.04~eV -- for the larger c-EC$^-$(EC)$_2$:Li$^+$
cluster.)  To compensate, 0.081~eV is added to the $\lambda$ when using
Eq.~\ref{lambda1} above.  This choice is made because adding an $e^-$ to the
most stable c-EC$^-$Li$^+$ geometry (Fig.~\ref{fig1}c) and then applying
geometry optimization lead to barrierless breaking of a
C$_{\rm C}$-O$_{\rm E}$ bond.  This finding underscores the fragility of
this bond and dovetails with fast CO evolution observed in AIMD
simulations.\cite{pccp}  However, estimating $\lambda$ in Eq.~\ref{marcus}
requires stable c-EC$^{2-}$:Li$^+$ and c-EC$^-$:Li$^+$ structures that
relax to each other after charge transfer.

\section{Crossover between one- and two-electron regimes}
\label{crossover}

The $\Phi$ and $\lambda$ values discussed above permit an estimate of
the crossover between one- and two-electron processes.  We neglect the
distinction between c-EC$^-$ and o-EC$^-$, which overly favors 1-$e^-$
processes, and assume no spatial homogeneity.  Following the definitions
in Table~\ref{table0},
\begin{eqnarray}
d[{\rm EC}]^-/dt &=& k_e [{\rm EC}] - k_e' [{\rm EC}^-] - k_1 [ {\rm EC}^-]^2 ;
	\label{1etmp}	\\
d[{\rm EC}^{2-}]/dt &=& k_e' [{\rm EC}^-]  - k_2 [ {\rm EC}^{2-}] .
\end{eqnarray}
In Eq.~\ref{1etmp}, we have omitted 1-$e^-$ oligomerization initiation
reactions which consume o-EC$^-$ and intact EC, but exhibit
considerable barriers.\cite{gewirth}  At steady state, 
\begin{eqnarray}
[{\rm EC}^{2-}] &=& k_e' [{\rm EC}^-] /k_2 ;  \\
{\rm [EC}^-{\rm ]}  &=& \large\{ -k_e' + (k_e'^2 + 4 k_1 k_e [{\rm EC}])^{1/2} 
	\large\} / 2 k_1 . \label{quad}
\end{eqnarray}
An upper limit of $k_1$, governed by diffusion, is $k_1 = 4\pi D a_o$.  Here
$D$ is the EC$^-$ diffusion constant and $a_o$ is the reaction radius.  We
adopt the rough estimates of $a_o$=2.5\AA\, and
$D$=10$^{-7}$cm$^2$/s, yielding $k_1$=10$^8$/($M$s).  In arriving
at this value, the radical recombination barrier to form BDC has been set
to zero.  If we had used the 0.15~eV barrier predicted by Bedrov {\it et al.},
the 2-$e^-$ regime will be even more prominent. If the 1-$e^-$ product is taken
to be EDC with a 0.61~eV formation barrier,\cite{bedrov} not BDC, $k_1$
becomes so slow that the 1-$e^-$ regime almost vanishes.

\begin{figure}
\centerline{\hbox{ \includegraphics[width=4.00in]{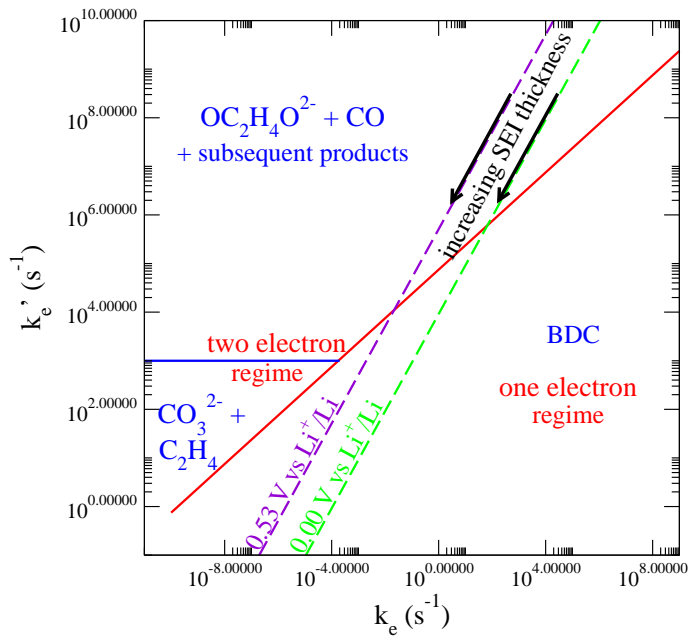} }}
\caption[]
{\label{fig4} \noindent
Different SEI formation regimes, assuming steady state reactions and
a homogeneous reaction zone.  The $x$- and $y$-axes are the 1-$e^-$
and 2-$e^-$ tunneling rates.  Green and violet dashed lines represent applied
potentials of 0.0 and 0.53~V versus Li$^+$/Li(s).  OC$_2$H$_4$O$^{2-}$ is not
the final product.  Oligomers are not explicitly considered but are secondary
products in the OC$_2$H$_4$O$^{2-}$ region and may be present at low k$_e$
rate in the ``BDC'' region.  We have not extended the voltage above 0.53~V,
needed to reach the CO$_3^{2-}$ region in this simple estimate.  In the
presence of electrode surfaces, CO$_3^{2-}$ formation rate can be much
faster (see text) and the CO$_3^{2-}$ regime expands.
}
\end{figure}

The red line in Fig.~\ref{fig4} denotes $k_e'^2$$=$4$k_1$$k_e$[EC].
From Eq.~\ref{quad}, 1-$e^-$ processes dominate below this line.
This more readily occurs at low one-electron tunneling rates $k_e$.
The main product is bimolecular recombination of o-EC$^-$
to form BDC.  If the electron transfer is extremely slow and [o-EC$^-$] is
low, o-EC$^-$ and intact EC may instead react to initiate 
oligomerization,\cite{gewirth,bedrov} or even proton exchange to form
CH$_3$CH$_2$OCO$_2^-$, despite the larger reaction barriers.

Above the red line, 2-$e^-$ processes dominate.  If the rate of $e^-$
transfer to c-EC$^-$ ($k_e'$) is slower than the c-EC$^-$ ring-opening rate
($k_3$, blue horizontal line), the second $e^-$ attacks ring-opened o-EC$^-$
to yield exclusively CO$_3^{2-}$ plus C$_2$H$_4$ products.  If however
$k_e'$$>$$k_3$, the second $e^-$ attacks c-EC$^-$ to form CO and doubly
deprotonated ethylene glycol (OC$_2$H$_2$O$^{2-}$).  Significant CO
gas has been detected during first charge,\cite{onuki,yoshida,ota1,ota2,shin} 
and carbon labeling techniques have demonstrated that much of the CO originates
from EC.\cite{onuki}  OC$_2$H$_4$O$^{2-}$ undergoes further reactions and may
yield products that can include C$_2$H$_4$ and CO$_3^{2-}$ (see the next
section).  Fig.~\ref{fig4} thus suggests multiple regimes with different SEI
compositions.  This appears consistent with the experimental interpretation
that SEI consists of an inner inorganic layer and an outer polymeric/organic
layer.\cite{review} On Cu electrode surfaces, the outer layer has been
demonstrated to be porous/spongy and can be penetrated by Li$^+$ and
counterions.\cite{harris} It has been suggested that the inner, inorganic SEI
layer consists of Li$_2$CO$_3$, although this has been disputed.\cite{edstrom}

$k_e$ and $k_e'$ are functions of distinct $V_o$'s (Eq.~\ref{marcus}) which
depend on the frontier orbitals of EC and c-EC$^-$, respectively.  Neglecting
this small orbital dependence, $V_o$ depends mainly on the tunneling barrier
through the nascent SEI film, which is a function of SEI thickness.  Then the
ratio $(k_e'/k_e)$ depends on the $\lambda$ and $\Delta G^o$ associated
with 1- and 2-$e^-$ additions, not SEI thickness.  Assuming a constant
applied voltage equal to that of Li$^+$/Li(s), and using the values estimated
in Sec.~\ref{reduction} and Sec.~\ref{reorg} for EC and c-EC$^-$ reduction,
$k_e'/k_e$$\sim$9$\times$10$^3$ is predicted.  {\it Thus 2-$e^-$ reduction
can be much faster than 1-$e^-$ reduction on a per EC basis.}  The green
dashed line depicts this constant $k_e'/k_e$ ratio.  The system progresses
along this line at constant voltage as the SEI thickens.  The purple dashed
line represent an 0.53~V applied voltage vs.~Li$^+$/Li(s).  It is
shifted to the left as applied voltage increases because 1-e$^-$ processes
exhibit a less favorable $\Phi$ and benefit more from lower voltages.  
As the predicted $\Phi$=0.53~V for 1-e$^-$ EC reduction (Table~\ref{table1}),
compared to the experimental $\sim$0.8~V, we have not depicted $(k_e'/k_e)$
for higher voltages needed in this estimate to reach the CO$_3^{2-}$+C$_2$H$_4$
region.  However, in the presence of electrode surfaces, the barrier of
C$_{\rm E}$-O$_{\rm E}$ bond-breaking can be drastically reduced (Fig.~3 \&~6
of Ref.~\onlinecite{ald}), subsequent CO$_3^{2-}$ production (Eq.~\ref{old2e})
can be much faster, and the CO$_3^{2-}$ region in Fig.~\ref{fig4} will expand.

The predicted boundaries are approximate due to the homogeneous reaction zone
assumption.  In reality, $k_e$, $k_e'$, and [c-EC$^-$] all develop spatial
profiles that tail off away from the electrode/electrolyte interface.  By
omitting electrodes, we may have overestimated the c-EC$^{2-}$
formation rate because $e^-$ tunneling to EC$^-$ will only be effective
near the electrode region while o-EC$^-$ recombination can occur anywhere.
Alternatively, we may have underestimated the c-EC$^{2-}$ formation rate
if EC$^-$ are strongly bound to the electrode surface instead of diffusing
freely in solution.  Finally, the boundaries will be blurred
by inhomogeneities present in battery electrode particles.

\section{Ultimate fate of 2-$e^-$ products}

\begin{figure}
\centerline{\hbox{ (a) \includegraphics[width=2.00in]{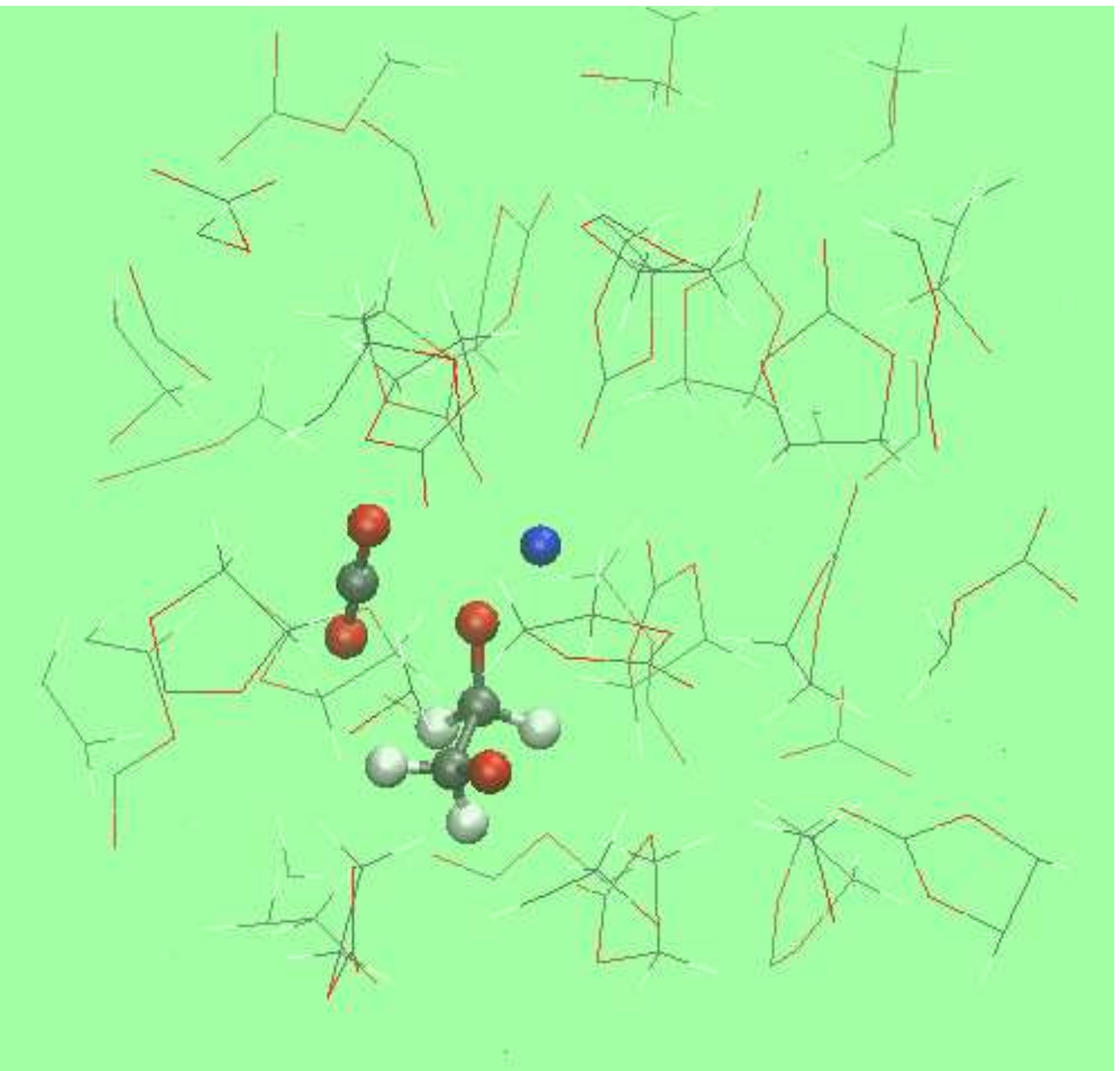}
                   (b) \includegraphics[width=2.00in]{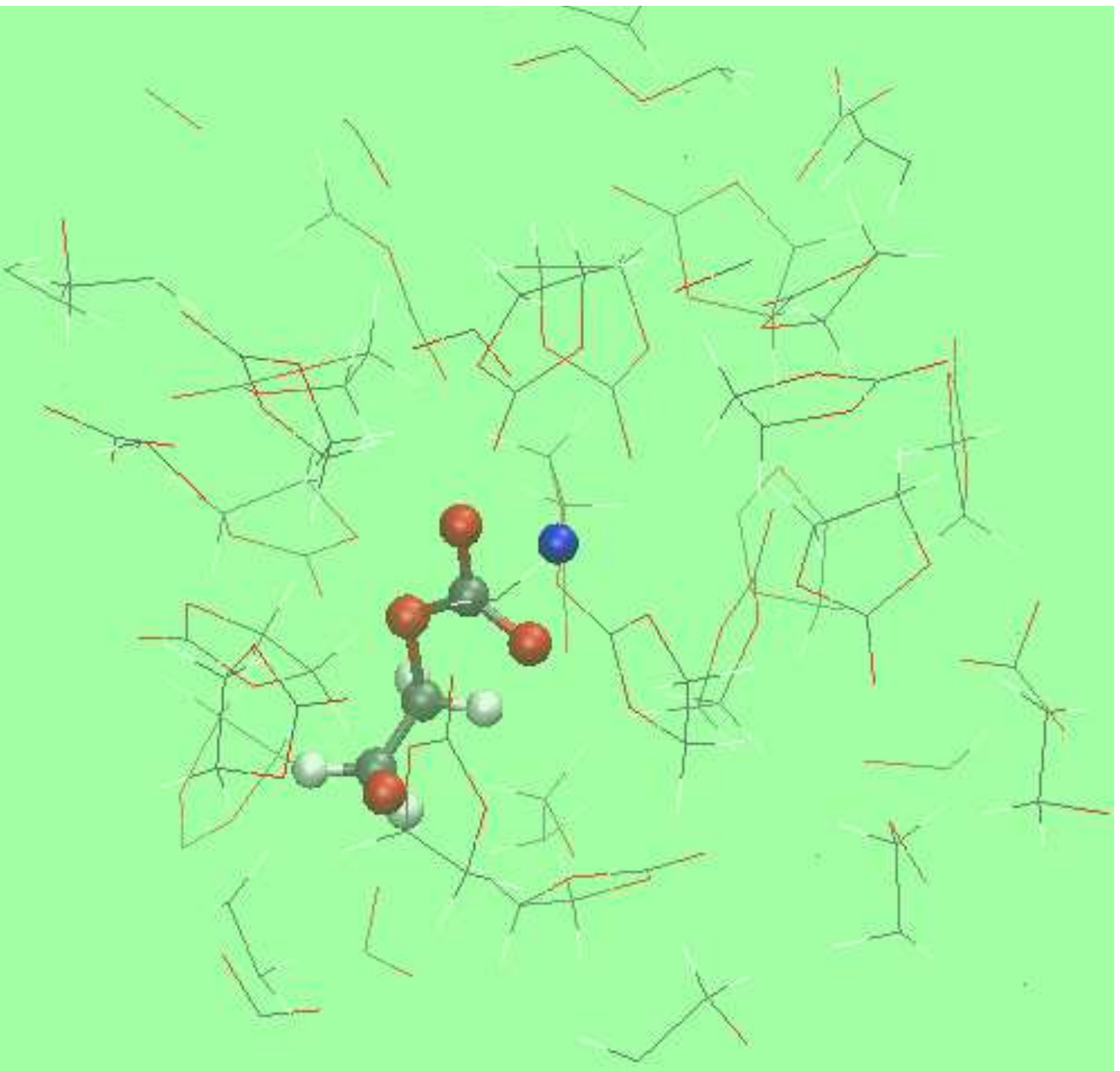} }}
\centerline{\hbox{ (c) \includegraphics[width=2.00in]{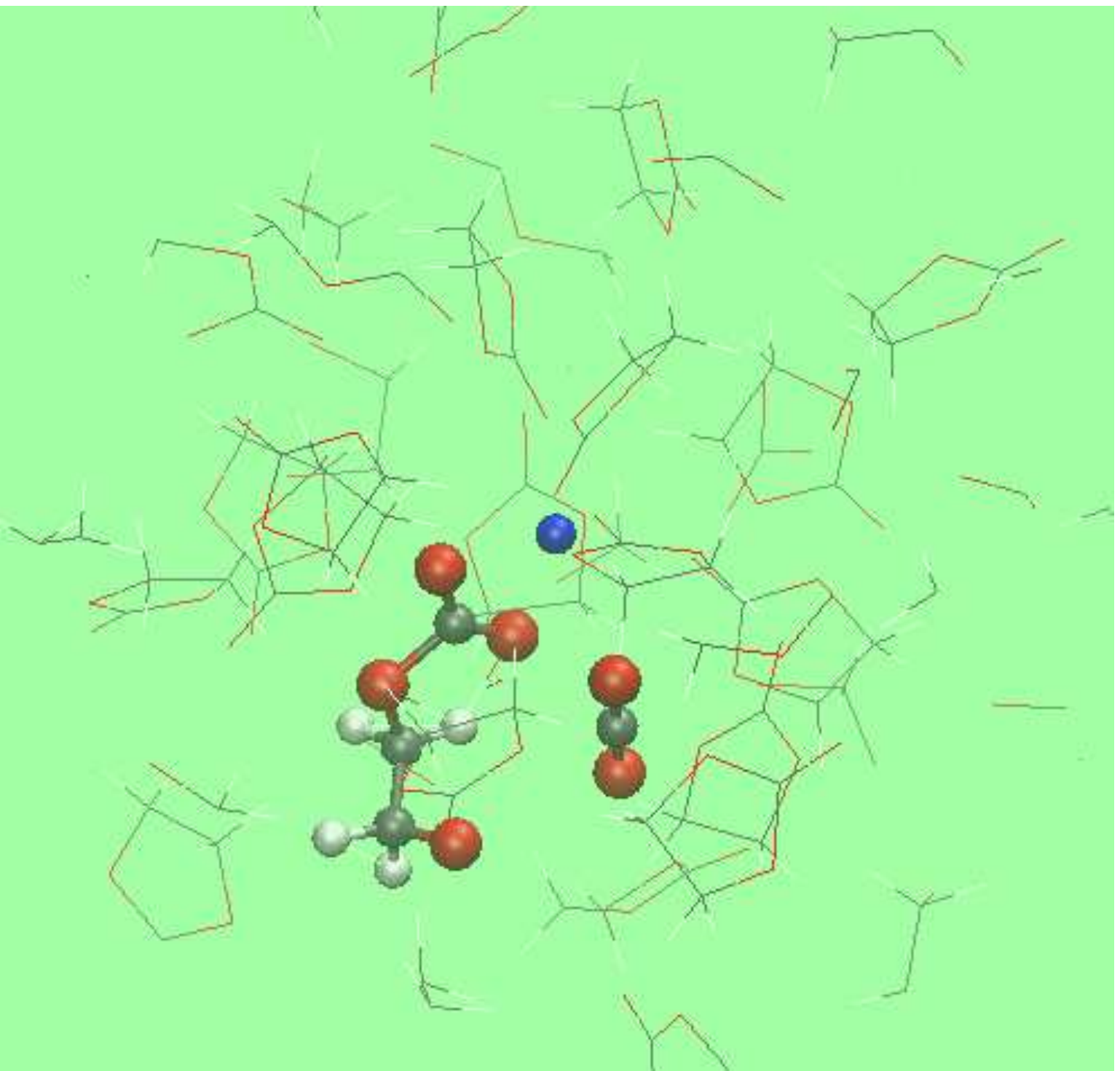}
                   (d) \includegraphics[width=2.00in]{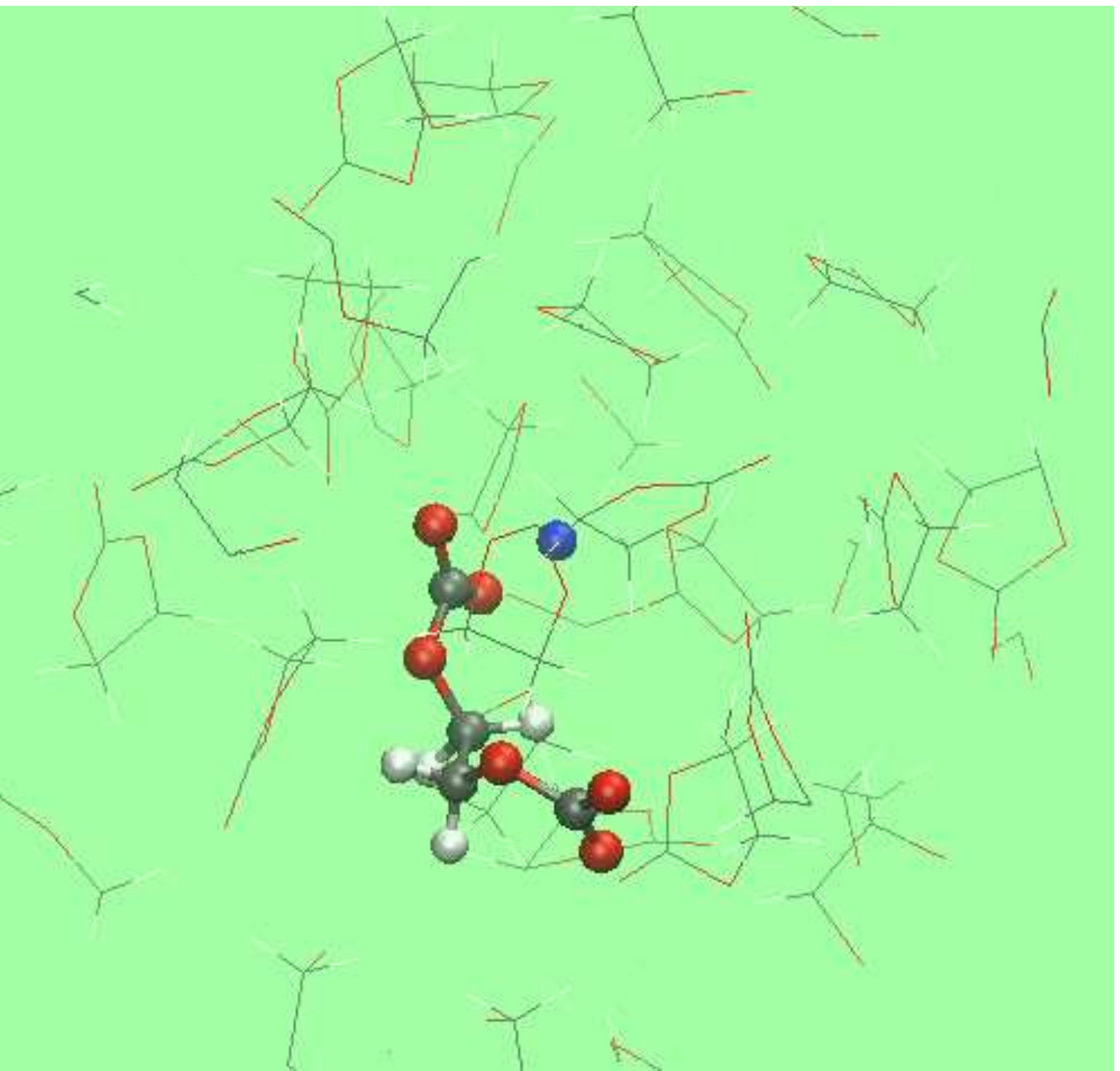} }}
\caption[]
{\label{fig5} \noindent
OC$_2$H$_4$O$^{2-}$ + 2 CO$_2$ $\rightarrow$ EDC, from AIMD
simulations.\cite{ald}
(a) AIMD snapshot of OC$_2$H$_4$O$^{2-}$ in liquid EC
    next to a CO$_2$ placed 3~\AA\, from it.
(b) Within 2 seconds, nucleophilic attack of CO$_2$ spontaneously
    occurs to form a semi-carbonate (OC$_2$H$_4$OCO$_2^{2-}$).
(c) Another CO$_2$ is artificially place close to the other end of
    the semicarbonate.
(d) Within 2 seconds, nucleophilic attack of the other CO$_2$ 
    occurs to form EDC (O$_2$COC$_2$H$_4$OCO$_2^{2-}$).  Spectactor
    EC are depicted as stick figures.
}
\end{figure}

OC$_2$H$_4$O$^{2-}$ has not been reported as part of the SEI.  Unlike
more stable semicarbonates (ROCO$_2^-$), it is very reactive and its
RO$^-$ termini readily attack other species.\cite{lee2000}  One possible
termination product for OC$_2$H$_4$O$^{2-}$ is in fact EDC which has been
assigned the main SEI product in FTIR experiments.\cite{edc} In AIMD
simulations, we have observed barrierless attack of OC$_2$H$_4$O$^{2-}$
on two CO$_2$ to form EDC (Fig.~\ref{fig5}).  This appears viable in full
cell experiments where large amount of CO$_2$ is released, much of it
originating from EC (presumably from reactions on the cathode).\cite{onuki}
In fact, this mechanism
is closely related to the laboratory EDC synthesis pathway which starts from
ethylene glycol and Li$_2$CO$_3$.\cite{edc}  However, this EDC formation
mechanism cannot be universal.  CO$_2$ evolution from the cathode does not
occur until a voltage of as much as 5~V vs.~Li$^+$/Li, or unless the cathode
contains Ni.\cite{novak1999}  Furthermore, in some graphitic carbon SEI
experiments, Li counter electrodes are applied.  CO$_2$ evolution from Li metal
is not expected.  Note that oxalic acid formation via combination of reduced
CO$_2^-$ radical anions will compete with mechanisms that involve
CO$_2$.\cite{tarascon}

\begin{figure}
\centerline{\hbox{ (a) \includegraphics[width=2.00in]{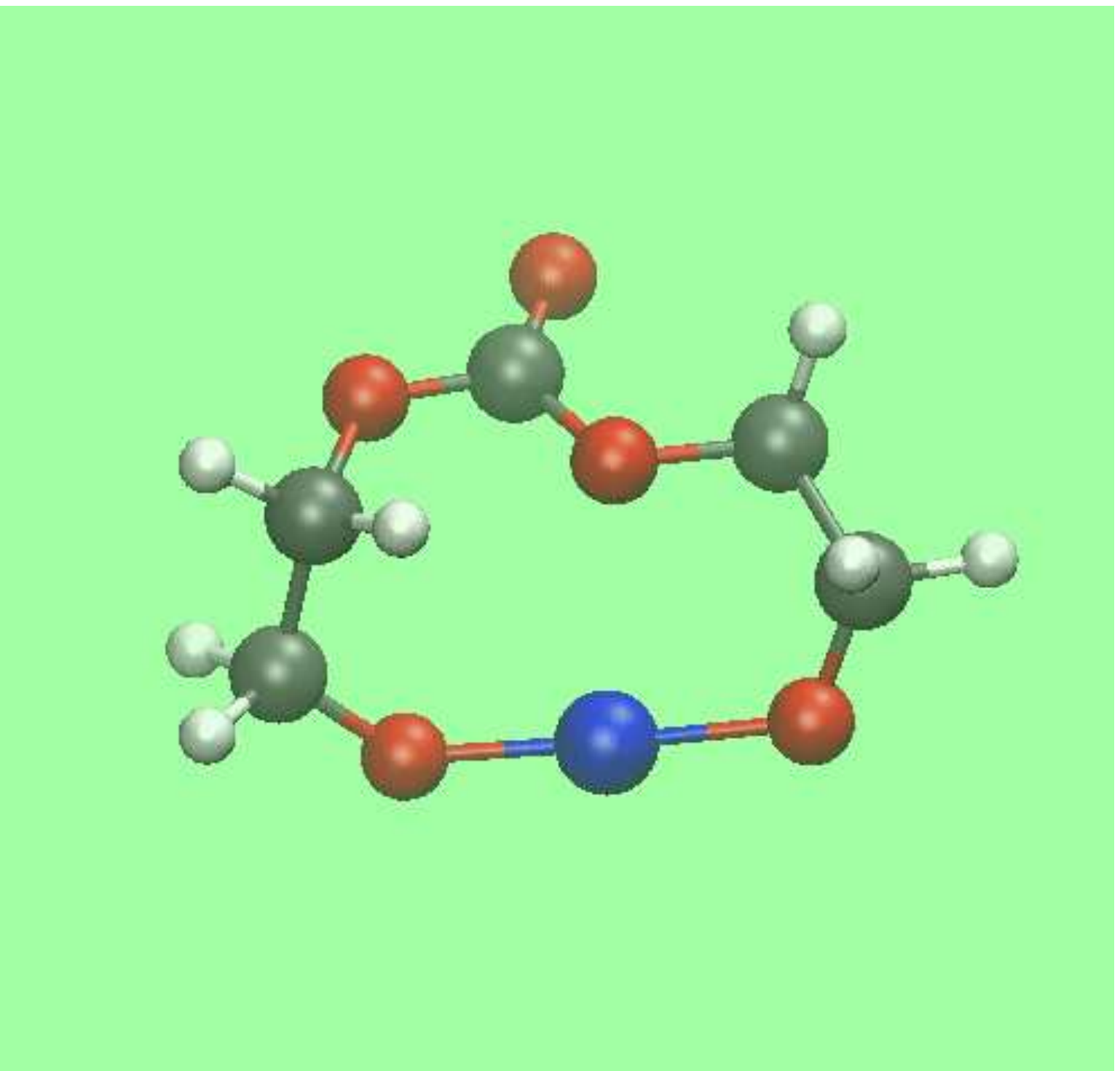}
                   (b) \includegraphics[width=2.00in]{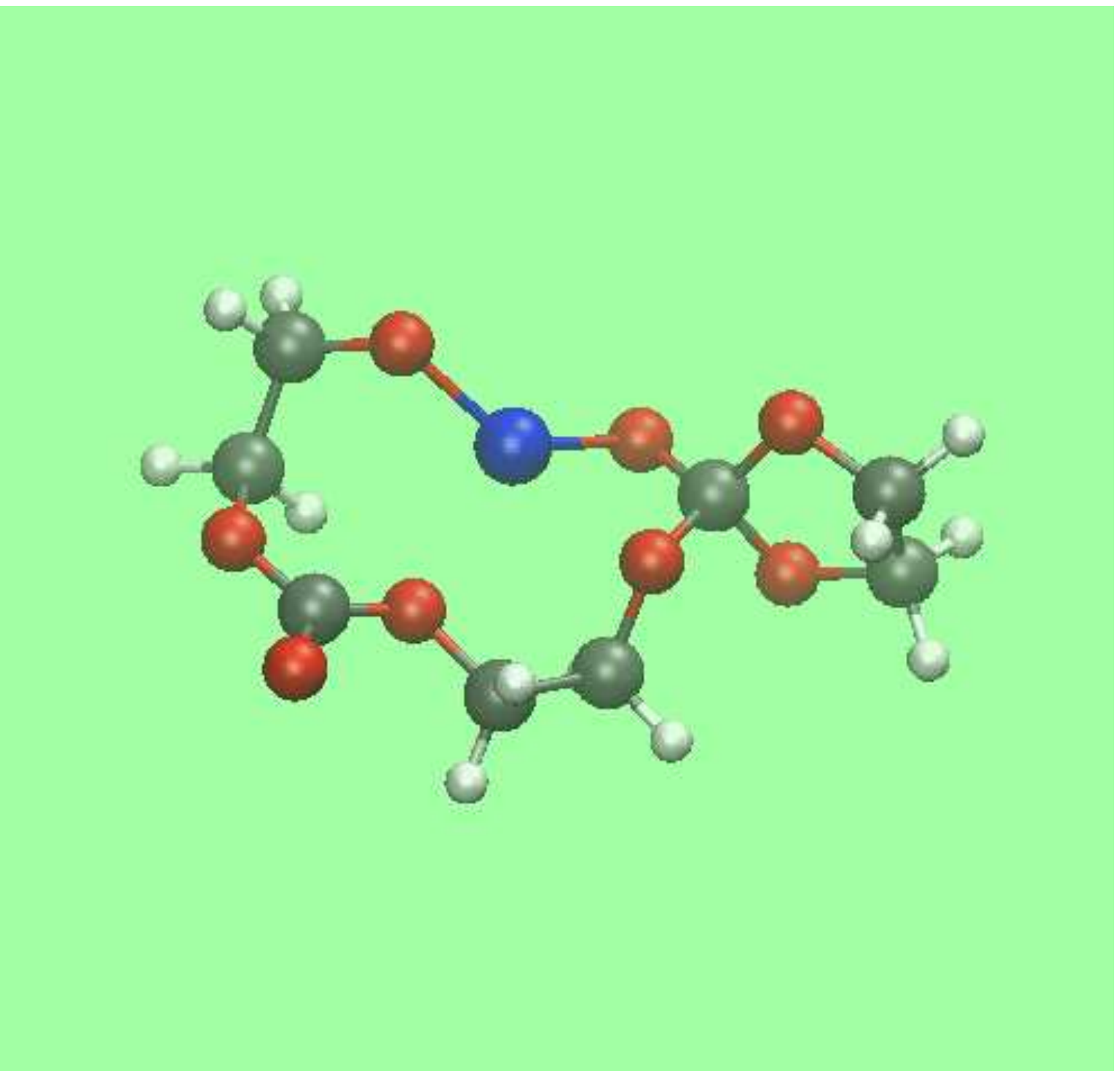} }}
\centerline{\hbox{ (a) \includegraphics[width=2.00in]{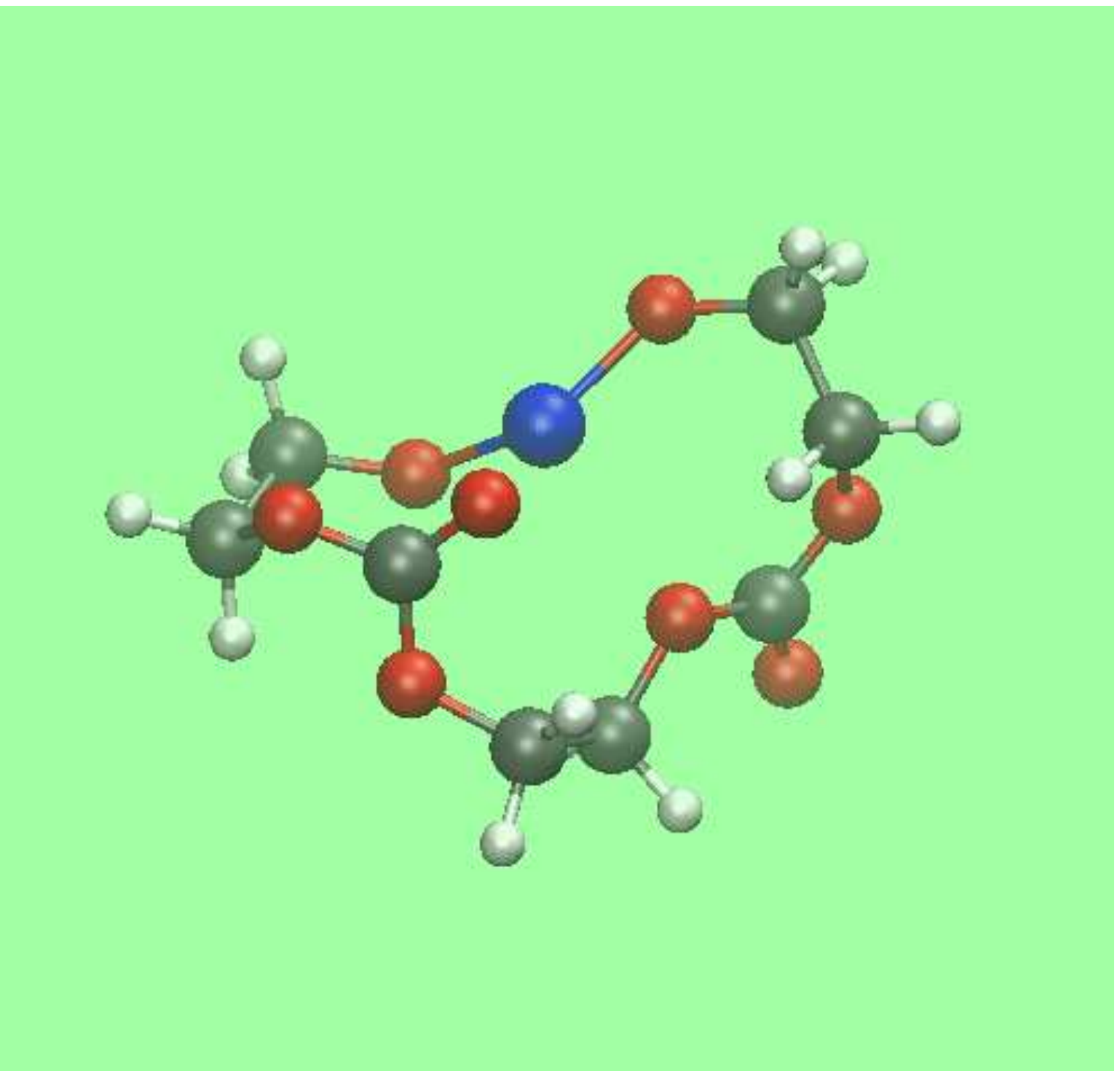}
                   (b) \includegraphics[width=2.00in]{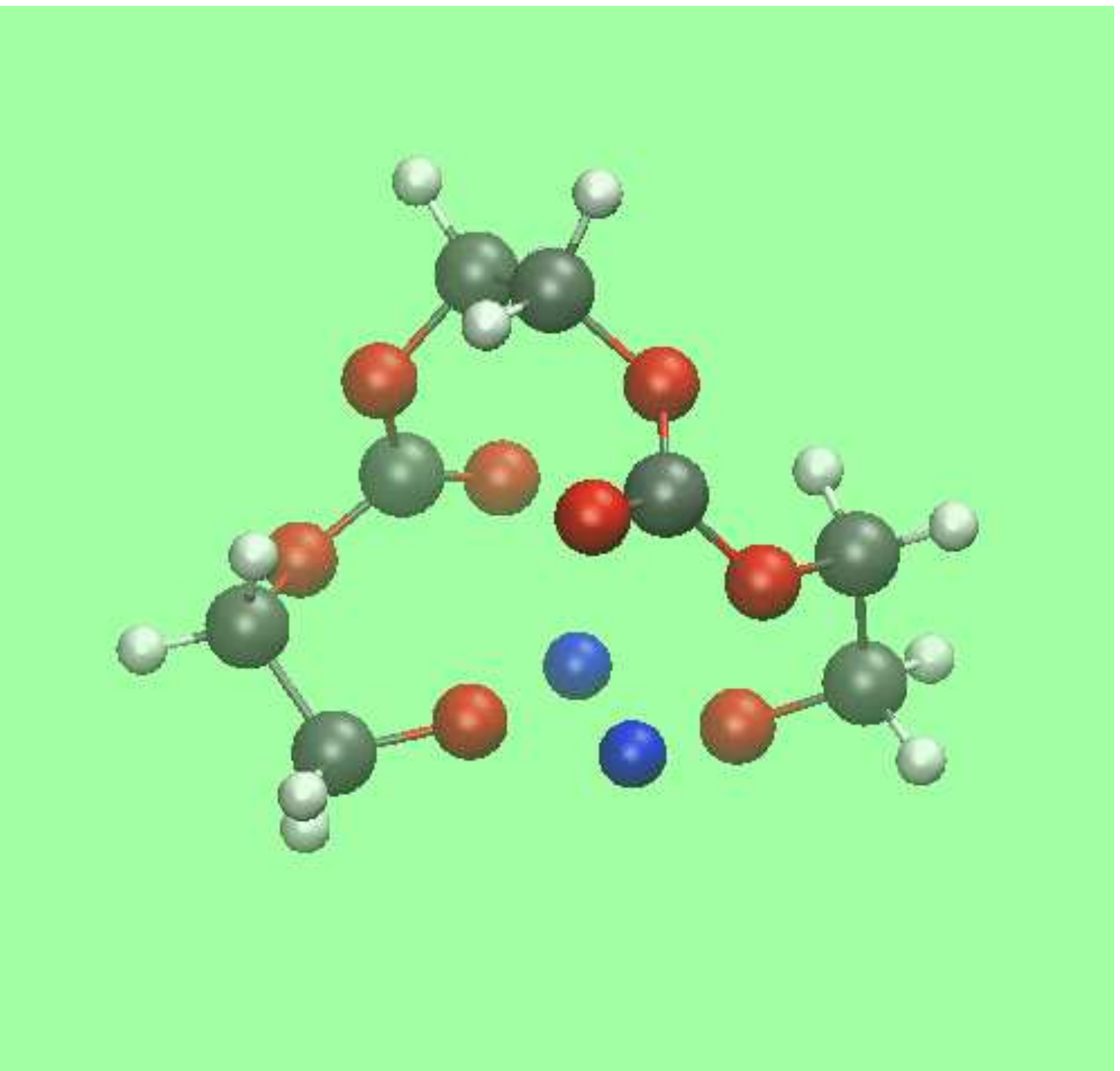} }}
\caption[]
{\label{fig6} \noindent
(a) OC$_2$H$_4$OCO$_2$C$_2$H$_4$O$^{2-}$:Li$^+$ from
   nucleophilic attack of OC$_2$H$_4$O$^{2-}$ on EC.
(b) Attack on a second EC yields
    EC-OC$_2$H$_4$OCO$_2$C$_2$H$_4$O$^{2-}$:Li$^+$,
   with the EC C$_{\rm C}$ atom 4-coordinated.
(c) The chain configuration 
  O(C$_2$H$_4$OCO$_2$)$_2$C$_2$H$_4$O$^{2-}$:Li$^+$ with a
   broken EC C$_{\rm C}$-O$_{\rm E}$ bond is metastable compared to (b).
(d) Adding a second Li$^+$ stabilizes the chain configuration of (c).
All calculations are conducted at $\epsilon$=40.
}
\end{figure}

As EC are plentiful in the electrolyte, the more likely subsequent
reactions are polycarbonate initiation/propagation:
\begin{eqnarray}
{\rm OC}_2{\rm H}_4{\rm O}^{2-}:{\rm Li}^+ + {\rm EC} &\rightarrow &
 {\rm OC}_2{\rm H}_4{\rm OCO}_2{\rm C}_2{\rm H}_4{\rm O}^{2-}:{\rm Li}^+ 
	\label{eg-ec1} \\
{\rm OC}_2{\rm H}_4{\rm O-EC}^{2-}:{\rm Li}^+ + {\rm EC} &\rightarrow &
 {\rm EC-OC}_2{\rm H}_4{\rm OCO}_2{\rm C}_2{\rm H}_4{\rm O}^{2-}:{\rm Li}^+ 
	\label{eg-ec2} \\
{\rm OC}_2{\rm H}_4{\rm O-EC}^{2-}:({\rm Li}^+)_2 + {\rm EC} &\rightarrow &
 {\rm O(C}_2{\rm H}_4{\rm OCO}_2{\rm )}_2{\rm C}_2{\rm H}_4{\rm O}^{2-}:({\rm Li}^+)_2 \label{eg-ec3}
\end{eqnarray}
The general predictions of our MP2 geometry optimization calculations are as
follows (Fig.~\ref{fig6}).  The terminal R-O$^-$ groups of OC$_2$H$_4$O$^{2-}$
readily attacks the C$_{\rm C}$ atom of intact EC, 
such that the target EC C$_{\rm C}$ atom becomes coordinated to
four different RO-.  Similar reactions have been suggested for alkoxide-intiated
attack on EC, with the alkoxide coming from DMC.\cite{tarascon,sasaki}
These C$_{\rm C}$-(OR)$_4$ motifs appear stable under some conditions (e.g.,
a dearth of coordinating Li$^+$ ions, Fig.~\ref{fig6}b).  Under other
conditions they spontaneously undergoes a ring-opening reaction by breaking
a C$_{\rm C}$-O$_{\rm E}$ bond (Fig.~\ref{fig6}a,c).  Related, KOH-induced
EC ring-opening and polymerization reactions have been reported.\cite{lee2000}

The polycarbonate chains in Fig.~\ref{fig6} contain reactive RO$^-$ termini
and can consume more intact EC molecules via repeated nucleophilic attacks.
The first product (Fig.~\ref{fig6}a) incurs barriers less than 0.2~eV.
As the chain lengthens, reaction barriers should increase and exothermicities
decrease.  Various chain-capping mechanisms have been proposed.\cite{lee2000}
DMC, often present in battery electrolyte but neglected in this work, may also
react with terminal RO$^-$ to yield -OCO$_2$CH$_3$ end groups.  Even if
that is the case, the interior of these chains still contains R-OCO$_2$-R'
linkages.  Like similar motifs in EC, DMC, and other simple organic carbonates,
they should also be susceptible to electrochemical reduction at low voltages
to possibly yield CO, CO$_3^{2-}$/C$_2$H$_4$, oligomers, and even EDC.  
In fact, Fig.~\ref{fig6}d clearly contains the EDC motif; EDC may be released
if two electrons are added and two C$_2$H$_4$O$^-$ radicals are elminated.
This suggests that 2-$e^-$ mechanisms can lead to multistep reactions and
a wide variety of products.\cite{ogumi}  Polymeric SEI components, especially
those with polyethylene oxide or ester signatures, have been reported on
graphitic anode surfaces.\cite{gewirth,tarascon}  Polycarbonates themselves
have been found on cathode surfaces\cite{pec_cathode} but not on anodes,
presumbly because they undergo further electrochemical reduction.

\section{Conclusions and Outlook}

In conclusion, we have reviewed recent theoretical predictions on
ethylene carbonate (EC) decomposition mechanisms in the bulk liquid
region, in the absence of electrodes or counterions.  Supplementary
calculations are performed to incorporate recent computational
insights about two-electron attacks on EC\cite{pccp,bal11,ald} into
the cluster-based framework first used by Balbuena and coworkers.\cite{bal01}

Our quantum chemistry (MP2) calculations show that reduction potentials
($\Phi$) rank in the order o-EC$^-$$>$c-EC$^-$$>$EC.  This implies that two
electron processes are favored and should be considered in interpretations of
SEI products.  Reorganization energies ($\lambda$) also favor two-electron
over one-electron attacks.  The cross-over between one- and two-electron
routes is estimated within a steady state, spatially homogeneous reaction zone
approximation.  Even though it is speculative and underestimates 
C$_2$H$_4$/CO$_3^{2-}$ products, it is valuable starting point for further
studies.

When electron tunneling is fast, our MP2 cluster-based product predictions
are consistent with AIMD simulations of electrode/electrolyte
interfaces.\cite{pccp,bal11,ald}  The initial two-electron mechanism products
are CO and doubly deprotonated ethylene glycol (OC$_2$H$_4$O$^{2-}$).  The
predicted CO release dovetails with some previous experimental
measurements\cite{ota1,ota2,onuki,yoshida,shin} and mechanistic
assignments.\cite{ota1,ota2,onuki,yoshida,shin,marom}  The ultimate fate
of OC$_2$H$_4$O$^{2-}$ is undetermined but may include CO, CO$_3^{2-}$,
C$_2$H$_4$, ethylene dicarbonate (if CO$_2$ is present), and even oligomers.
The decomposition of vinylene carbonate radical anions should also be
re-examined in light of the predicted weakness of the C$_{\rm C}$-O bond.   

When tunneling of the second $e^-$ from the electrode is slow compared to
EC$^-$ ring-opening, two $e^-$ attack on EC produces C$_2$H$_4$ and CO$_3^{2-}$
instead.  This route has been more widely associated with two-electron attacks.
It is more exothermic but, in the absence of electrodes, exhibits a higher
barrier than the CO-route.  The energetics\cite{bal01,bedrov} and kinetic
barriers\cite{bedrov} reported in previous electronic calculations are
consistent with the fact that one-electron reactions should yield butylene
dicarbonate (BDC), not etheylene dicarbonate (EDC) --- at least in the
absence of counterions and/or electrode surfaces.  AIMD simulations and our
new MP2 calculations suggest that EDC can arise from 2-$e^-$, multistep
processes.

Most calculations are conducted under idealized settings, neglecting electrodes,
anions, cosolvents, spatial inhomogeneity, and other experimental details.
True SEI formation mechanisms are undoubtedly even more complex.  The
theoretical calculations reviewed and newly performed herein only highlight
kinetically or thermodynamically favorable mechanisms and products which
{\it can} occur.  To prove that such reactions actually take place in battery
settings requires measurements.  To that caveat, it must be added that most
mechanisms proposed in the experimental literature derive from considerations
of final SEI product distributions.  Such extrapolation can be unreliable if
multistep reactions take place.  Electronic calculations can predict accurate
reaction barriers and help interogate previously proposed pathways.  Theory
also provides useful guidelines and structural motifs for intepretations of
future SEI-related experiments and can yield useful input for multiscale
calculations.\cite{oakridge,kee}  Regarding future theoretical studies, the
specific role of counter ions, cosolvents, electrode surfaces, the ultimate
fate of fast two-electron reaction products, and more rigorous ways to deal with
electron tunneling between the anode and electrolyte are clearly urgently
needed.  Accounting for spatial distributions of reduced species will yield
more accurate 1-$e^-$/2-$e^-$ crossover estimates.

\section*{Acknowledgement}
 
We thank Leah Appelhans, Yue Qi, Oleg Borodin, John Sullivan, Nick Hudak, Kevin
Zavadil, Rick Muller, and David Rogers.  Sandia National Laboratories is a
multiprogram laboratory managed and operated by Sandia Corporation, a wholly
owned subsidiary of Lockheed Martin Corporation, for the U.S.~Deparment of
Energy's National Nuclear Security Administration under contract
DE-AC04-94AL85000.  This research used resources of the National Energy
Research Scientific Computing Center, which is supported by the Office of
Science of the U.S. Department of Energy under Contract No. DE-AC02-05CH11231.
KL is partially supported by the Department of Energy under Award Number
DE-PI0000012.

\end{document}